\documentclass{article}
\usepackage{cite}
\usepackage{amsmath,amssymb,amsfonts}
\usepackage{algorithmic}
\usepackage{graphicx}
\usepackage{float}
\usepackage[normalem]{ulem}
\usepackage[keeplastbox]{flushend}
\usepackage{balance}
\usepackage[utf8]{inputenc}

\title{Avoiding collisions at any (low) cost: ADS-B like position broadcast for UAVs}
\author{Franco Minucci, Evgenii Vinogradov and Sofie Pollin.}
\date{ESAT-Telemic, KU Leuven, March 2020}

\begin{document}

\maketitle
\begin{abstract}
Unmanned Aerial Vehicles (UAVs), a.k.a. drones, are increasingly used for different tasks.  
With more drones in the sky, the risk of accidents rises, sparkling the need for conflict management solutions.
Aircraft use a system called Automatic Dependent System-Broadcast (ADS-B) to continuously broadcast their position and speed but this system is not suitable for small drones because of its cost, complexity and capacity limitations. 
 
Broadband technologies such as Wi-Fi beacons are more suited for such dense scenarios, and they also offer the benefit of wide availability and low cost. 
The main challenges for Wi-Fi are (a) the multi-channel nature of the technology makes transmitter and receiver coordination difficult, and (b) standard chipsets are not designed for frequent broadcast transmission and reception. 
In this paper, we propose and analyze a multi-channel position broadcast solution that is robust against jamming and achieves a reliable location update within 125~ms. 
In addition, we implement the protocol on inexpensive embedded Wi-Fi modules and analyze the hardware limitations of such devices. 
Our conclusions are that even on the simplest Wi-Fi chipsets, our protocol can be implemented to achieve a realistic location broadcast solution that still perfectly mimics simulation and analytical results on the lab bench and still can achieve approximately 4~message/s throughput at a distance of 900~m on flying UAVs.

\end{abstract}
\section{Introduction}
\label{sec:introduction}
With UAVs (a.k.a. drones) becoming more and more important for a broad number of applications \cite{mozaffari,liu2014review,6616093,sallouha2018energy,vinogradov2018tutorial}, the traffic in our skies increases drastically \cite{pwc_report}. Because the vertical airspace in which UAVs are allowed to fly is narrow both because of legal and technical limitations, the probability of accidents due to collisions grows rapidly with increasing UAV density. 
In order to avoid accidents, each drone needs to be aware of the position and speed of the surrounding drones so that it can keep a safe separation distance with the other UAVs. Many initiatives are moving in this direction such as ICAROUS (based on ADS-B)\cite{consiglio2019sense}, Open Drone ID\cite{opendroneid} and the PercEvite project \cite{percevite} which the research presented here is part of.

\subsection{Problem Statement}

Imagine two drones travelling in opposite direction at 10~m/s, which is an average speed for typical quad-copters. 
Their relative speed would be 20~m/s. 
If the drones are 1~km apart, they are going to collide within 50~s.
This means that in less than one minute the two UAVs have to detect each other, and perform a maneuver to avoid the crash. 
A more rigorous study of the safety separation distance between drones is given in \cite{Weinert_mit}: the authors explicitly compute well-clear distance recommendations for different kind of drones, how to assess the risk of collision and prove their claims with simulation results.

It is obvious that the task of drones coordination consists of two main parts: i) obtaining the UAV positions and ii) sharing this information between UAVs. Both subtasks have own performance requirements.

\subsubsection{Positioning Update}
Since UAVs move quite fast, they also need to share their position frequently to enable tracking the correct, possibly adjusted, trajectory. 
With a typical GPS system, this means location updates should be received at least once per second. 
The update frequency of 1~Hz is a de facto standard more than a hard constraint. 
Its legacy comes from a set of requirements for car navigation systems. 
Modern inexpensive GPS modules can give location updates at frequencies up to 5~Hz while the fastest commercial modules update their position at rates up to 100~Hz. 
The only constraints are processing power, communication speed and location accuracy which degrades with the update rate.\cite{gps_misra}.

GPS positioning, however, is often enhanced by other sensors and technologies to have a better accuracy and faster updates\cite{locating_the_nodes}.
This is typically achieved by collecting data from ultrasound sensors, lidars, cameras, compasses and accelerometers and performing so called sensor fusion using Kalman filters. In this work, we consider that a UAV can obtain its positions with any required update frequency.

\subsubsection{Wireless technologies for the coordinate exchange}
\label{wireless_tech}
Wireless communication plays a vital role in Air Traffic Management (ATM) systems: the key technology in manned aviation is Automatic Dependent Surveillance - Broadcast (ADS-B). It is a surveillance technology in which an aircraft determines its position via satellite navigation and broadcasts it every 500 ms. The information can be received by air traffic control ground stations and by other aircraft to provide situational awareness and allow self separation.

Future Unmanned Traffic Management (UTM) systems will require a similar wireless technology adapted for the UAVs' needs and capabilities. It is obvious for the academic community and other major actors in the field of airspace management (e.g. NASA) that ADS-B will not be used for the small UAV traffic management.
As it was pointed out by the authors of \cite{consiglio2019sense,guterres2017ads}, using ADS-B in a large fleet of UAVs could overload and impair the existing air traffic management systems compromising the safety of manned aviation.

This understanding resulted in several coordinate exchange solutions based on APRS\cite{ads-b_like_utm}, LoRa\cite{Vinogradov2019WirelessCF,sharpe} and others (a more detailed overview is given in Section~\ref{Sec:SoA}). 
All these solutions do not require establishing a connection between the drones since this procedure can take too much time.
These solutions offer several kilometers range, however, their performance (in particular, the delay between two consecutive messages) is not sufficient for Conflict Management procedures especially at close distance (for a more detailed discussion refer to Section~\ref{sec:CM} and \cite{Vinogradov2019WirelessCF}). Moreover, broadcasting messages over an unnecessarily long distance is more detrimental than beneficial due to the increased interference. Additionally, the aforementioned solutions require installation of an additional payload.

Considering the listed disadvantages, we suggest to use a Wi-Fi based solution which has the potential to offer a high coordinate exchange rate while re-using the equipment that is commonly installed on small UAVs.

To answer the question whether it is possible to use Wi-Fi beacons as replacement for ADS-B messages, it is important to address the following points:
\begin{itemize}
    \item Working distance: How far can two UAVs detect each other?
    \item Reception rate (or throughput): At which rate can two UAVs exchange their position?
    \item Connection overhead: Is it possible to exchange positional information avoiding any connection overhead?
    \item Number of aircraft in the same air space: How many drones can broadcast their position in the same area before the radio collisions become the limiting factor for safety? 
\end{itemize}

The goal of this research is to study and conceive a system that can operate in a similar fashion to ADS-B using the limited Wi-Fi hardware typically used for telemetry in small commercial drones.
We aim to propose a protocol 
guaranteeing the performance needed for executing the air conflict management (CM) procedures (see Section~\ref{sec:CM} for more details).

\subsection{State of the art}\label{Sec:SoA}

In literature, there are many studies addressing UTM and which wireless technology might be the best to adopt for this purpose. Let us describe the most representative works.

\subsubsection{ADS-B-based solutions}
Authors of \cite{guterres2017ads} identified the problems of using standard ADS-B. 
Namely, since the transmission power for ADS-B on normal airplanes is very high and the air traffic is quite intense, the ADS-B receiver gets quickly saturated by messages coming from aeroplanes that do not pose any real threat to the UAV nor does the UAV represent any threat for aircraft which are tens of km away. In \cite{strohmeier2014realities}, the authors explicitly mention how ADS-B receivers can suffer for channel congestion in high traffic conditions or be "blinded" by transmitters that are in close proximity (<50~km) due to the high transmission power. 

Another important issue is the power consumption: standard ADS-B transponder uses 200~W transmit power. Even small transponders can go up to 20~W power \cite{ping1090}, which is extremely high for a small battery powered drone. 

Moreover, ADS-B transceivers which are specialized for UAVs are much more expensive than a small UAV. The ping RX, used in most of the studies, is just a receiver and it already costs approximately 250~\$, which is already the price of a small UAV. 
A full fledged ADS-B transponder for UAV, like the ping1090 by uAvionix\cite{ping1090}, costs 2000~\$. The price of these devices is mostly dictated by the qualification and certification process more than by their components, which means that it does not follow an economy of scale. 
These class of price is affordable for bigger UAVs but totally out of proportion for the small UAVs category (<5~kg) targeted by our system.

In \cite{consiglio2019sense}, the authors studied the performance of the 
Reduced Power (RP) ADS-B. 
In their study, the authors considered a version of ADS-B which uses only between 400~mW and 1.3~W instead of 200~W transmission power. 
With RP~ADS-B, the throughput drops to one message every two or three seconds which is not enough for safe conflict management. 


\subsubsection{Solutions based on other popular wireless technologies}
FLARM \cite{8514548,flarm} is a proprietary technology aiming to propose a solution alternative to ADS-B.
The protocol is proprietary, the patent is owned by FLARM g.m.b.h., and many details are kept confidential. 
FLARM uses a message rate of 3~msg/s with an effective bit rate of 50~kbps (100~kbps with Manchester encoding).
The max transmit power is 25~mW.

LoRa proves itself as a good candidate for exchanging messages at large distances. Authors of \cite{Vinogradov2019WirelessCF} report 5 and 30 seconds update delay for UAVs at 1 and 2 kilometers distance, respectively. The weakness of LoRa is the constraint on its duty cycle which limits the maximum message rate (minimum 5 s between two consecutive messages). It is possible to extend the communication range for the cost of increasing the delay between messages. Note that LoRa-based solutions might require the ground infrastructure but it is also possible to use it for the direct UAV-to-UAV communication as in \cite{Vinogradov2019WirelessCF}. 

Automatic Packet Reporting System (APRS) is a system designed in mid 80s. The data link layer follows the AX.25 protocol with a bit rate of 1200 bps.
Despite its simplicity and its widespread use especially in east Asia, it suffers from its age. However, APRS was used in \cite{ads-b_like_utm} and demonstrated the results similar to the ones of LoRa-based solution: 11 - 33 s update delay and a very long range (around 20 km).
The system needs a ground infrastructure to relay messages from one station to the other. It also suffers from radio collisions because it adopts quite long frames which are repeated multiple times to ensure their reception. 

\begin{table}[t]
    \centering
    \begin{tabular}{|l|r|c|l|}
    \hline
    \textbf{Technology} & \textbf{Range} & \textbf{Update rate} & \textbf{Max. Power} \\
    \hline
    Bluetooth LE & 50~m & 100~msg/s & 10~mW\\
    \hline
    Bluetooth & 100~m & 100~msg/s & 100~mW\\
    \hline
    RP ADS-B\cite{consiglio2019sense}& 1.2~km & 0.33 - 0.5~msg/s & 400~mW - 1.3~W\\ 
    \hline
    LoRa\cite{Vinogradov2019WirelessCF} & 10~km & 0.2~msg/s & 160~mW\\
    \hline
    FLARM\cite{8514548}  & 10~km & 0.33~msg/s & 25~mW\\
    \hline
    APRS\cite{ads-b_like_utm}& 20~km & 0.2~msg/s & 300~mW\\
    \hline
    ADS-B & 370~km & 2~msg/s & 200~W\\
    \hline
    \end{tabular}
    \caption{Comparison between alternative technologies that can be used for conflict management.}
    \label{tab:techs}
\end{table}

Table \ref{tab:techs} summarizes the characteristics of some of the common wireless technologies used or proposed in literature to perform the conflict management tasks.


\subsubsection{Wi-Fi-based solution}
We propose to broadcast positional information using Wi-Fi management frames, a.k.a beacons.
With Wi-Fi beacons, the proposed system manages to achieve a throughput of at least 2~msg/s in the worst communication conditions using only 17~dBm which corresponds to only 50~mW transmission power.

While normal Wi-Fi connections allow for higher data rates than broadcasting beacons, the disadvantages of standard Wi-Fi in this scenario are: 
\begin{itemize}
    \item The time to establish a connection might be too long \cite{8057164};
    \item The adaptive rate adaptation does not cope well with mobility \cite{Asadpour_gnd2air};
    \item An established connection may be an open door for hijacking the mission or stealing important data\cite{hacking_drone}\cite{drone_security}\cite{drone_hacking};
\end{itemize}

The congestion issue is less severe with Wi-Fi beacons compared to ADS-B since the limited transmission power makes it impossible to receive beacons from sources much further than one kilometer. Nonetheless, it is important to keep in mind possible interference by ground sources. Access points of office or home Wi-Fi networks \cite{van2015analysis} and other systems (such as Bluetooth or even drones remote controllers) are all examples of possible interference sources.

In our previous work \cite{8734208}, we performed and initial investigation of using Wi-Fi beacons to spread positional information between UAVs. 
The proposed scheme overcomes the main disadvantages of ADS-B and of traditional Wi-Fi communications for this specific application. It can be easily deployed on small commercial drones by a firmware update. 
However, in \cite{8734208}, we did not deeply analyze the protocol performance outside of our lab environment, its coexistence with command and control link, and its resilience against radio collisions.


\subsection{Main contributions and structure of the article}

The research presented in this article proposes a system based on Wi-Fi beacons to spread positional information. The solution incorporates also the simultaneous use of the Wi-Fi hardware for networking/telemetry, which is its primary purpose, other than broadcasting and receiving beacons.
The proposed Wi-Fi based system is 
a pragmatic solution for small UAVs (<5~kg).

This paper contributions can be summarized as: 
\begin{itemize}
    \item We propose a UAV coordinate exchange protocol based on encoding the positional information into the Wi-Fi Service Set Identifier (SSID) field;
    \item We design a mathematical and simulation models for the performance assessment;
    \item We present an open-source implementation of the protocol;
    \item We identify and underline important hardware constraints;
    \item We assess the protocol performance (dependency on the protocol parameters, range, throughput, and number of UAVs) via experiments in the lab scenario as well as in a real-world scenario.  
\end{itemize}

This paper is organized as follows: Section \ref{sec:background} provides some background on conflict resolution and Wi-Fi beacons; 
Section \ref{sec:system_model} introduces the proposed Wi-Fi location update protocol with both a statistical and a behavioral model;
Section \ref{sec:protocol_analysis} is about simulating the model for a range of scenarios in order to optimize the design parameters;
In Section \ref{sec:experimental_results}, an experimental validation of the model is performed on Wi-Fi hardware and flying drones.
Finally, Section \ref{sec:wifi_modules} contains important lessons learned during the experimental validation and some in depth information about the embedded Wi-Fi modules used for our research.
\section{Background}
\label{sec:background}

This section introduces the background on the conflict management and the Wi-Fi technology.

\subsection{Air conflict management}\label{sec:CM}
\begin{figure}
    \centering
    \includegraphics[width=0.8\linewidth]{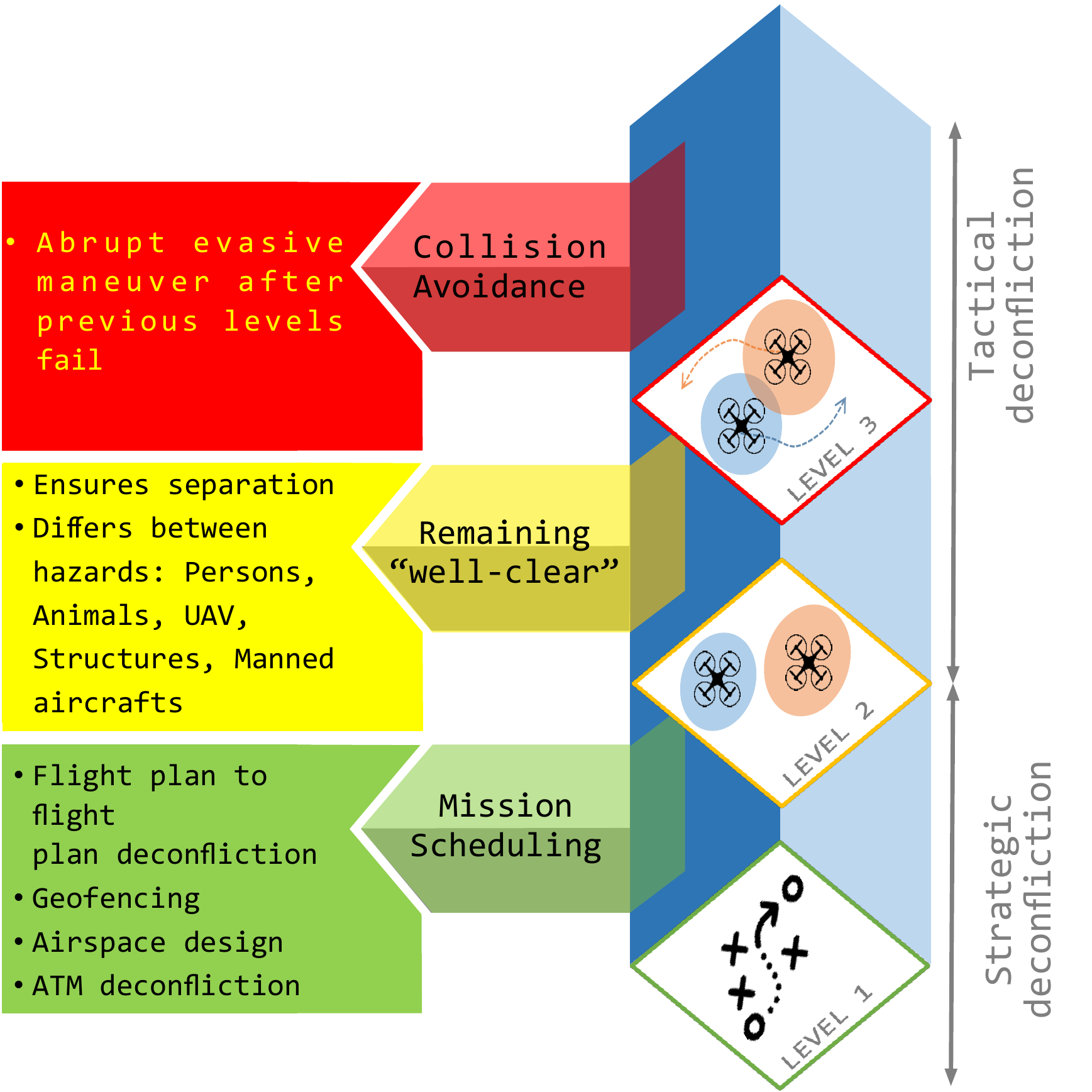}
    \caption{Relation between conflict management layers and collision timeline}
    \label{fig:cm_levels}
\end{figure}
When two or more aircraft have to share a common air-space there is a possibility that their trajectories intersect causing accidents.
These events are called conflicts and thus the act of foreseeing and preventing them is known as conflict management. 
The more aircrafts occupying the same volume of space, the higher the probability of conflicts. 
This is well known to the aviation industry which along the years put together a strict set of rules that pilots and ground personnel have to follow during each flight.

Conflict management, as explained in \cite{Vinogradov2019WirelessCF}, can be either strategic, which means planning way-points and trajectories in advance; or tactical, which means dealing with possible conflicts arising during the flight itself by negotiating with the other aircraft in proximity and the control towers and maneuvering to prevent possible accidents.
Figure \ref{fig:cm_levels} illustrates the different layers of conflict management.
Beware that tactical and strategic management are complementary and both should be in place to guarantee safety.

Strategic conflict management consist of assigning flight plans to each aeroplane in order to minimize the number of air encounters. 
However, in some cases, aircraft can only pass close to each other. 
When this happens, they have to keep a safety distance between each other called well-clear separation. 
When it is not possible to keep well-clear separation, we enter the realm of collision avoidance, which means avoiding a collision at all costs.

Broadcasting UAVs positions is a prerequisite for both strategic and tactical conflict management because drones do not have a pilot on board which can look out of the windows and maneuver according to what he sees.
Although camera based collision avoidance has been studied extensively \cite{sedagath_pisheh, mcguire_decroon}, it has the drawback of limited range, high computational power and one or multiple cameras to be effective.

Broadcasting the position and speed of each UAV opens up the possibility to perform well-clear separation before the UAVs enter the area of collision avoidance by adopting specific algorithms like Optimal Reciprocal Collision Avoidance\cite{orca}\cite{clearpath}. 

\subsection{Beacons}
\begin{figure}
    \centering
    \includegraphics[width=0.9\linewidth]{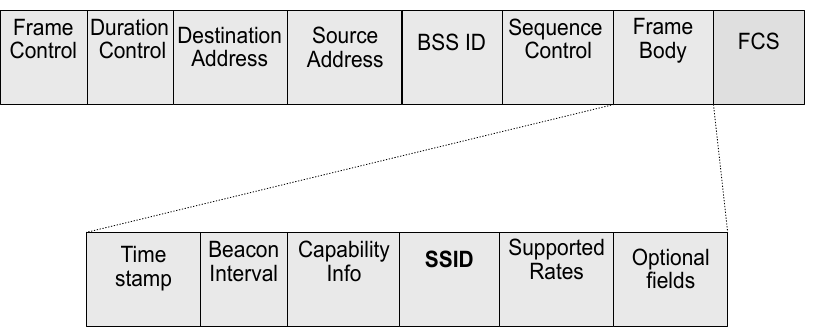}
    \caption{Beacon Frame structure: the information can be encoded in the SSID field which can host 32 bytes without constraints.}
    \label{fig:beacon}
\end{figure}

For the reasons described in the section\ref{wireless_tech}, Wi-Fi beacons seem to be a good candidate to absolve this task. 
Since beacons use the lowest bitrate allowed by the selected Wi-Fi standard, they can propagate very far, even in the range of a kilometer. 
Beacons are normally sent every 102.5~ms giving enough margin to receive at least one beacon per second even if many are lost. They also do not require any acknowledgement nor any connection establishment to be received so there is no connection overhead in the communication.
The positional information can be easily encoded and embedded in the SSID field (see Figure \ref{fig:beacon}).

In principle, it is also possible to embed information in other fields of the beacon frame but then care must be taken to still keep the frame valid. Our analysis results does not depend on this implementation detail. 
Differently from the SSID, the other fields are interpreted by the WI-Fi stack and not by default reported to the user. 
This means that if collision avoidance information is placed in another field than the SSID, it may render the beacon invalid. 
In this case, the receiving Wi-Fi hardware will discard the message without reporting its content even though it is successfully received. 
\section{System Model}
\label{sec:system_model}
In this section we first introduce the protocol. Afterwards, we propose a model for it and a simple analysis framework. 

\subsection{Protocol}
\subsubsection{Channel selection}

One key challenge for Wi-Fi is the rendezvous protocol to ensure that the transmitter and receiver are working on the same channel at the same time with as little coordination as possible.
This requires state transitions between scan, broadcast and data communication modes which might take a long time, especially on low cost hardware.  
In our previous work\cite{8734208}, we investigated multiple strategies to exploit the Wi-Fi hardware for location broadcasting.
\begin{figure}
    \centering
    \includegraphics[width=1.0\linewidth]{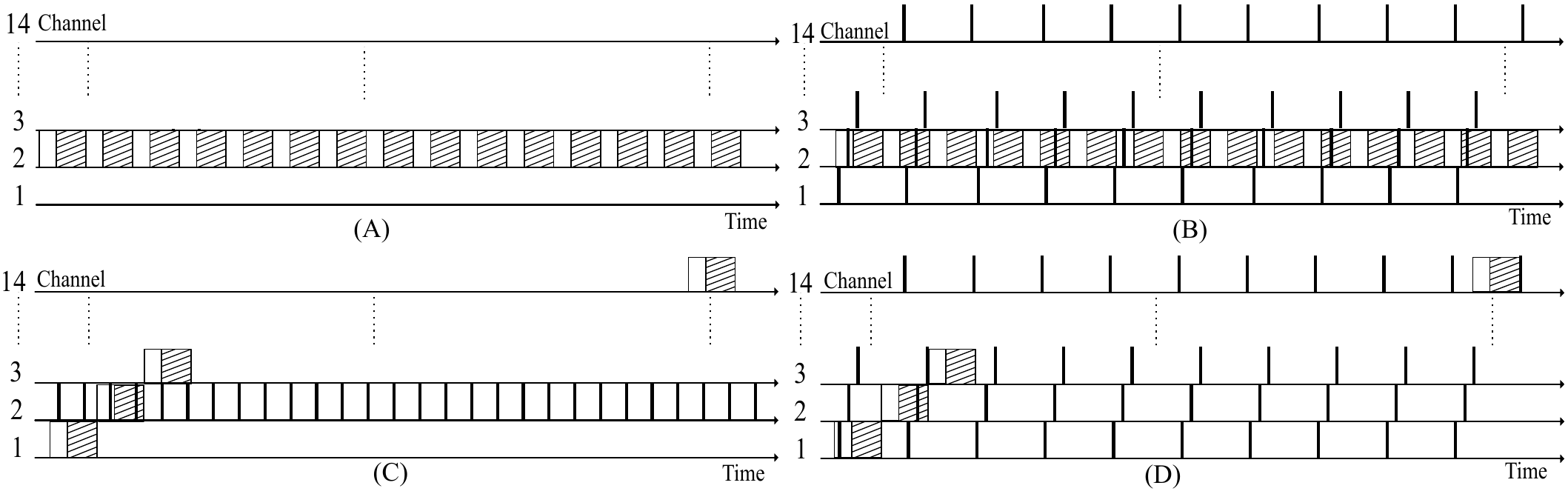}
    \caption{Broadcast/scan strategies: (A): broadcasts and scans use a single channel; (B)  broadcasts are on all channels, scans on a single channel, (C) broadcasts are on a single channel, scans are on all channels, (D) broadcasts and scans span all channels. Transmissions are indicated by vertical bars, scans by rectangles. The white part of scans is where the receiver is actually listening and the remaining part is the time used to process the incoming transmission.}
    \label{fig:strategies}
\end{figure}
In particular, we identified four possible modes of operation (See Figure \ref{fig:strategies}):
\begin{enumerate}
    \item[A] broadcasting and scanning on the same channel;
    \item[B] broadcasting on all the channels and scanning a single channel;
    \item[C] broadcasting on a single channel but scanning all the channels;
    \item[D] broadcasting on and scanning all the channels.
\end{enumerate}
The most effective strategy among the four resulted being Strategy B.
While strategy A is the fastest, it requires both transmitter and receiver to know in advance which channel is used for the collision avoidance broadcast. 
If the channel is congested, both need coordination to change it.
The main drawback of strategies C and D is that, scanning all channels is a long operation which can take more than two seconds and the received messages are only available at the end of it. 

Strategy B does not require any particular rendezvous algorithm and allows for continuous scanning of a single channel which can be chosen depending on traffic or propagation conditions.
Strategy B can also be modified to use only a subset of the available channels to reduce the impact on other Wi-Fi networks.

\subsubsection{Timing}
\begin{figure}[h]
    \centering
    \includegraphics[width=0.8\linewidth]{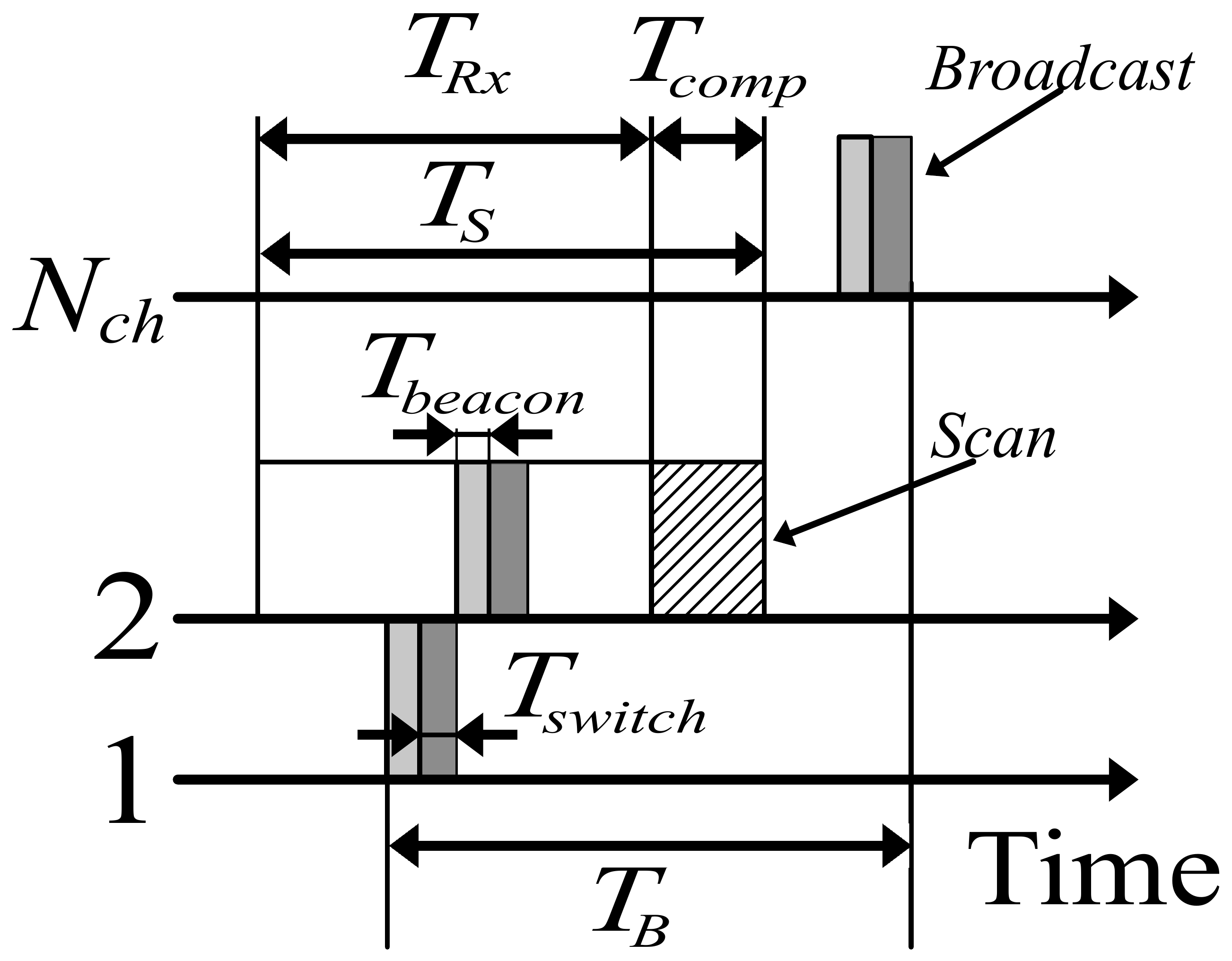}
    \caption{While in broadcast state, a drone broadcasts on all the available channels sequentially. Conversely, scans are always performed on the same channel. 
The result is that during a broadcast operation, each channel is only occupied for a time $T_{Beacon}$ and not for a time $T_{B}$.}
    \label{fig:TB_Tbeacon}
\end{figure}
The vast majority of inexpensive Wi-Fi modules have only one RF chain. This implies that the operations of transmitting and receiving messages cannot be performed simultaneously. In this light, the understanding of timing is vital for the protocol design. 

The duration of a beacon is indicated as $T_{beacon}$ but the duration of a multi-channel broadcast event is $T_{B}$. During $T_{B}$, multiple transmissions take place with a small interval in between in order to change channel. 
This means that, while the radio stays in broadcast state for a period of time $T_{B}$, it actually occupies each channel only for $T_{beacon}\times N_{ch}$. However, some time ($T_{switch}$ in Figure \ref{fig:TB_Tbeacon}) is also required to jump from one transmit channel to the next one. The relation between $T_B$ and $T_{beacon}$ is $T_B = N_{ch}\cdot (T_{beacon}+T_{switch})$ (see Fig.~\ref{fig:TB_Tbeacon}).

The duration of a scan operation by a radio in receive mode is indicated as $T_{S}$.
$T_S$ incorporates both the time in which the module is actively listening to the channel $T_{Rx}$ and the time spent to process the received transmissions and extract the packets $T_{comp}$. Thus $T_S = T_{Rx}+T_{comp}$.
This is illustrated in the timing diagram in Fig.~\ref{fig:TB_Tbeacon}.
It is worth noting that the $T_S$ may increase significantly when beacon traffic increases. 
However, when the number of received transmissions is low $T_{comp} \approx 0$ and $T_S \approx T_{Rx}$.

The increased duration of $T_S$ may effectively reduce the success rate since the additional time spent for processing is time in which the radio module is not listening for incoming transmissions. 
For low traffic, the processing time is so small that it can be effectively neglected. 
A more in depth analysis of this phenomenon is presented in Section \ref{sec:wifi_modules}.

With $T_{N}$, we then indicate the time dedicated to traditional Wi-Fi networking operations. 
A networking operation is time allocated to exchange telemetry or other data via Wi-Fi (e.g. video streaming or sensor readings).

\subsection{Behavioral model}
Consider a set of drones, each one equipped with a radio transceiver which can be in three possible states:
\begin{itemize}
    \item Broadcast with duration $T_B$: the node broadcasts positional information in beacon frames an all the Wi-Fi channels;
    \item Networking with duration $T_N$: the node is blind to beacons and is using the radio for other tasks like telemetry or video streaming on a single channel;
    \item Scan with duration $T_S$: the node listens for incoming beacons on the same channel used for telemetry.
     
\end{itemize}

At each instant in time, each module can be in any of the three mentioned states independently from the state of any other node composing the group of UAVs.

When a node is in scan state, it scans a single Wi-Fi channel for beacons and filters only the ones that contain positional information.
Different channels offer different performance and thus it is wise to implement different scan channel selection strategies depending on the UAV mission or environment. The design of those strategies is out of scope for this work.

We assume that networking operations do not collide with broadcasting because Wi-Fi uses Carrier Sense Multiple Access/Collision Avoidance which employs a listen-before-transmit mechanism\cite{bianchi_wifi}. We do study collisions between beacon frames. 

Some Wi-Fi modules can work in monitor mode and can receive multiple kind of packets at the same time (beacons, association, IP packets, etc...).
However, monitor mode is not standard in Wi-Fi adapters and therefore, it is not taken into account by our model.

\subsubsection{Probabilistic model}
We now model the protocol statistically, assuming the radio randomly selects one of the three states after completing the previous event with duratin $T_{B,S,N}$. 
The 
probabilities of being the broadcast state, scan state or network state are defined respectively as $P_{B}$, $P_{S}$ and $P_{N}$, respectively. In other words, they describe the share of time spent in the corresponding state.

The state selection probabilities are indicated as $\rho_{B}$, $\rho_{S}$ and $\rho_{N}$. The model assumes that the probability of selecting a state is independent from the previous state. 

Since the duration of each state once it starts is different, $P_{B,S,N}$ are different from their selection probabilities $\rho_{B,S,N}$.
To clarify, let us consider as an example a case in which no networking is involved; if $P_{B}=0.5$, $T_{B}=30~ms$, $P_{S}=0.5$ and $T_{S}=60~ms$, then for each scan operation there will be two broadcasts, which means that the selection probability for the broadcast state $\rho_B$ is twice as much as the transition probability for the scan state.
The selection probabilities are linked to the steady state probabilities as in:

\begin{eqnarray}
    P_B = \frac{\rho_B \cdot T_B}{\rho_B \cdot T_B+\rho_S \cdot T_S+\rho_N \cdot T_N} \label{eq:rB}\\
    P_S = \frac{\rho_S \cdot T_B}{\rho_B \cdot T_B+\rho_S \cdot T_S+\rho_N \cdot T_N} \label{eq:rS}\\
    P_N = \frac{\rho_N \cdot T_B}{\rho_B \cdot T_B+\rho_S \cdot T_S+\rho_N \cdot T_N} \label{eq:rN}
\end{eqnarray}
Solving these equations for $\rho_B$, $\rho_S$ and $\rho_N$ gives the selection probabilities.

The probability of observing the device actually broadcasting a beacon is:
\begin{equation}
    P_{beacon} = P_{B} \cdot T_{beacon}/T_{B}.
\end{equation}
When $k$ drones are in the system, which means more drones are close enough to be in range of the broadcasting drone and all can transmit,
the probability of a successful reception is:
\begin{equation}
P_{success}(k) = P_{S} \cdot P_{B} \cdot (1-P_{collision}(k)).
\end{equation}
$k$ is the number of potential transmitters.
The meaning of this equation is that the probability of success is defined as the probability that one drone is listening, another drone is broadcasting and the others are not broadcasting but rather scanning or networking.

The probability that, in a scenario with $k$ drones, given one drone is transmitting a beacon, at least another one tries to transmit is:
\begin{equation}
    P_{collision}(k) = 1-(1-P_{beacon})^{k-1}.
    \label{eq:P_collision_kdrones}
\end{equation}

%

In our analysis, we are interested in the probability to get a position update in a given window, and we assume here that 
$T_{W}=1$~s is a meaninful value. 
Given $T_W$, 

we can derive the average amount of events within an observation window as:
\begin{equation}
\overline{N_{x}} = P_{x} \cdot \frac{T_{W}}{T_{x}},
\end{equation}
where $x$ can be $S$, $N$ or $B$ which stand for Scan, Networking and Broadcast,  respectively.

From the number of broadcasts we can derive the average number of successful broadcasts during the observation window $T_{W}$:
\begin{equation}
\label{eq:n_succ}
    \overline{N}_{success}(k) = P_{S}P_{B}(1-P_{collision}(k)) \cdot \frac{T_{W}}{T_B}.
\end{equation}

\section{Protocol Analysis}
\label{sec:protocol_analysis}
\subsection{Behavioral Simulation}
In order to validate our system, we designed a behavioral simulation in which each node is modeled as a state machine. At each step of the simulation, a random number generator provides the input to determine which transition will be performed next.

Success probability, collision probability, and other statistical metrics are analyzed based on the simulation results.
The time step resolution for the simulation is 1~ms which is also the minimum time resolution reliably measurable on the physical device used for the experiments.

\subsection{Parameters Tuning}
Analyzing the results obtained with different parameters combinations allows for a much quicker evaluation of the multiple trade offs involved in designing such a system. Moreover, it helps to test the protocol in conditions that are difficult to achieve with the hardware itself (e.g. collisions).
The timing values used for the simulations are listed in Table \ref{tab:sim_timingl}. All the source code used to conduct all the experiments and simulations is available at the URL in \cite{percevite_repo}.
\begin{table}[ht]
\caption{Simulation Parameters. The timing values are chosen according to our measurements on the low-cost embedded Wi-Fi modules. These values are device dependent and can be changed if different Wi-Fi modules are used. Using these values is important to compare simulation results with the experimental results.}
    \centering
    \begin{tabular}{|l|c|}
        \hline
        \textbf{Parameter} & \textbf{Value} \\ \hline
        $T_{beacon}$ & 1~ms\\ \hline
        $T_{B}$ & 30~ms\\ \hline
        $T_S$ & 60~ms\\ \hline
        $T_N$& 100~ms\\ \hline
        Time resolution & 1~ms\\ \hline
        Number of drones & $\geq 10$\\ \hline
        Number of state transitions & $10^6$ \\ \hline
    \end{tabular}
    
    \label{tab:sim_timingl}
\end{table}
\begin{figure}[t]
    \centering
    \includegraphics[width=0.9\linewidth]{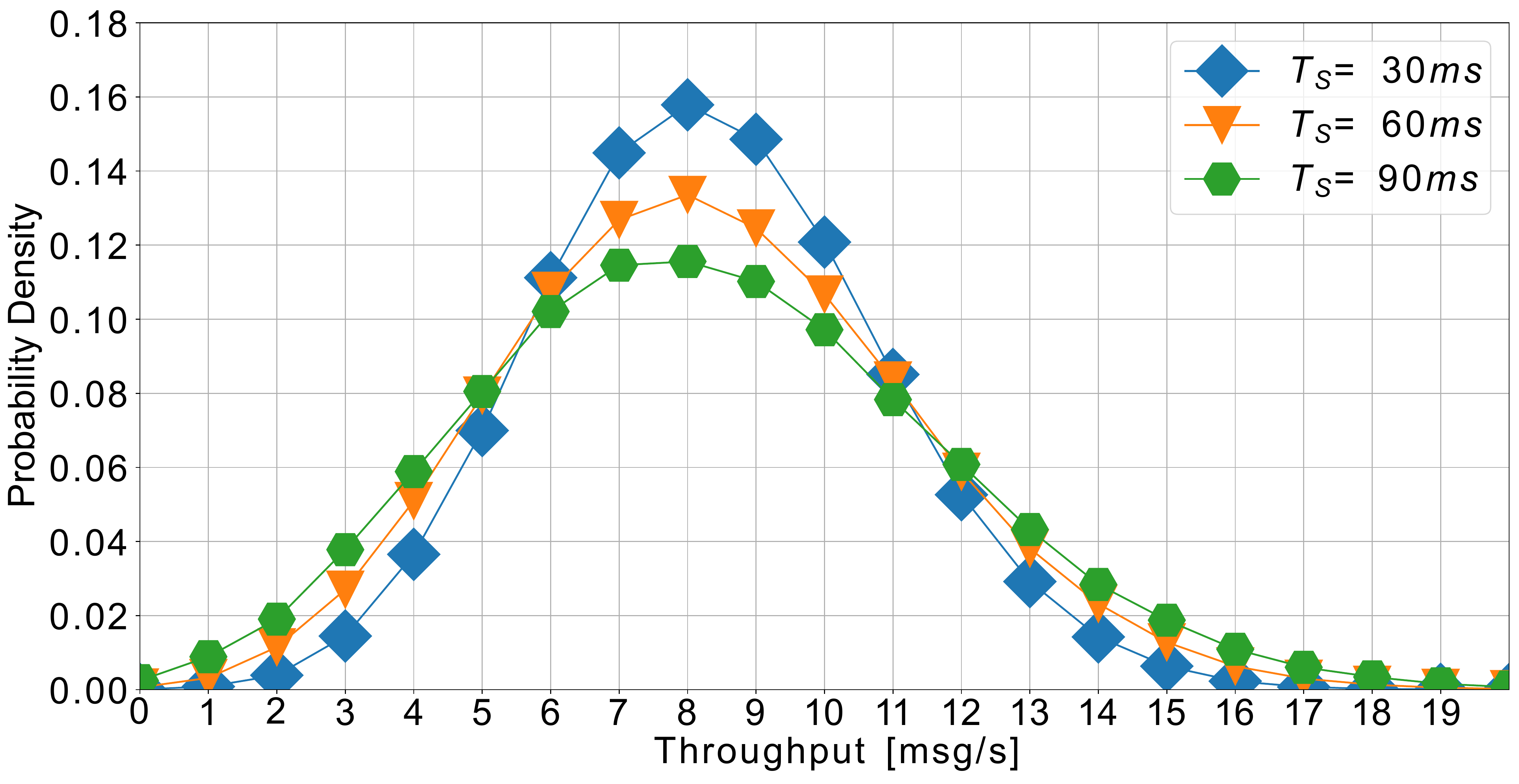}
    \caption{Probability density of receiving k messages in a second. The peak is the average throughput. Keeping $P_S = P_B = 0.5$ and $T_B=30$~ms, the impact of $T_S$ on the was analysed. As expected, a shorter scan burst gives a lower spread on the number of received messages, although it proved difficult to achieve on the real hardware. However, an intermediate value of 60 ms still offers good performance while being more hardware friendly. }
    \label{fig:TS_tuning}
\end{figure}

Figure \ref{fig:TS_tuning} illustrates how the Broadcast - Scan throughput is affected by changing the duration of a scan operation. Shorter and more frequent scans result is a narrower distribution and a higher probability of keeping the average throughput. However, increasing $T_S$ does not produce any important change in the protocol performance.
Because of our hardware, we have decided to chose $T_S=60$~ms. A more detailed explanation about the hardware limitations on $T_S$ is given in section \ref{sec:wifi_modules}. 
\begin{figure}[t]
    \centering
    \includegraphics[width=0.9\linewidth]{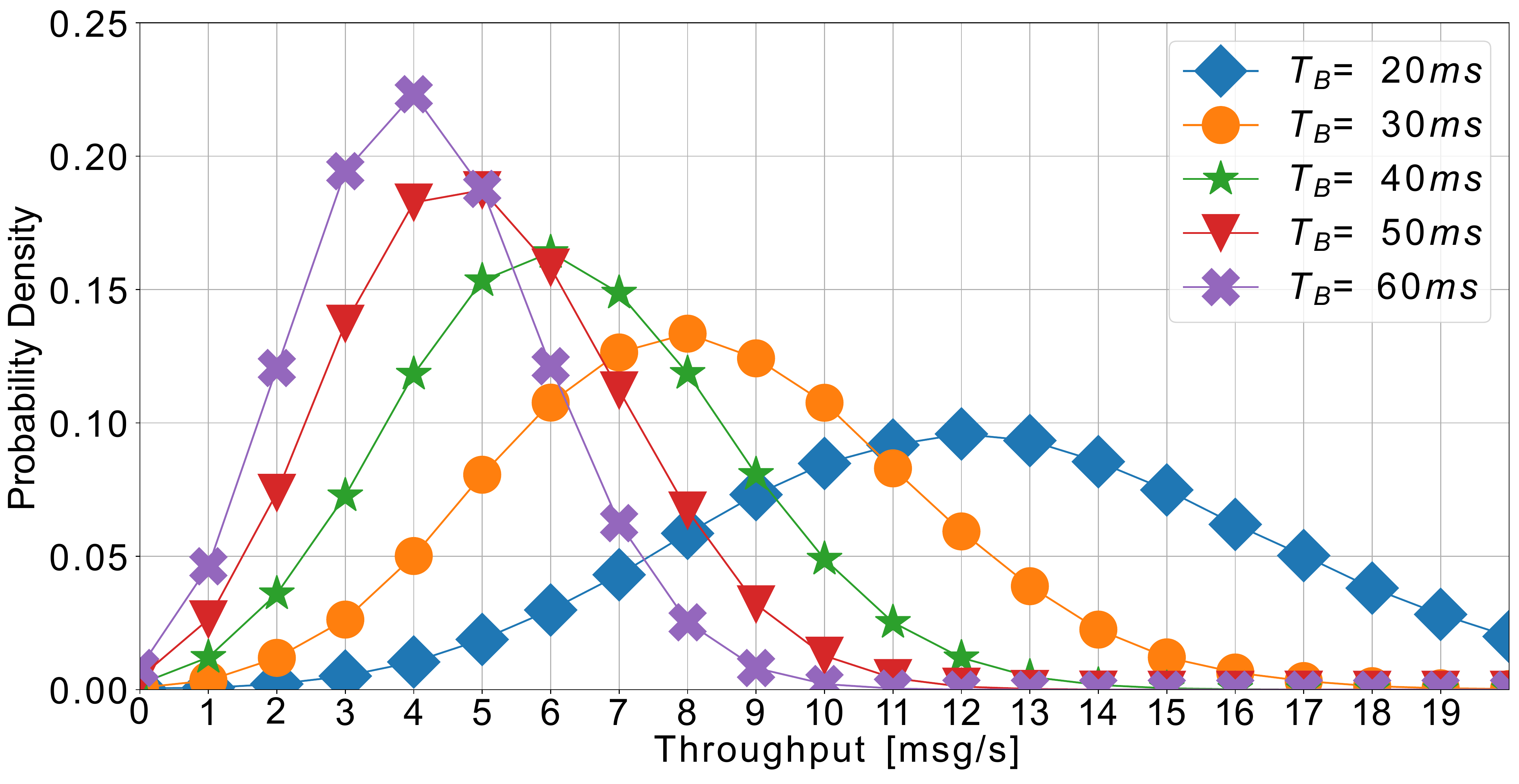}
    \caption{Probability density of receiving k messages in a second. The peak is the average throughput. Keeping $P_S=P_B=0.5$ and $T_S=60$~ms, the PDF was evaluated for different values of $T_B$. The trend shows increasing Throughput for shorter $T_B$. The selection of $T_B$ proves to have a much grater impact on the throughput than the selection of $T_S$.}
    \label{fig:TB_tuning}
\end{figure}

As shown in Figure \ref{fig:TB_tuning}, the value of $T_B$ has a larger impact on the broadcast - scan throughput. 
Minimizing $T_B$ is important to increase the throughput but the limitation of the real hardware must be taken into account when deciding a value. 
The lower limit to $T_B$ is given by the duration of a beacon plus the setup time of the hardware itself.
\begin{figure}[t]
    \centering
    \includegraphics[width=0.9\linewidth]{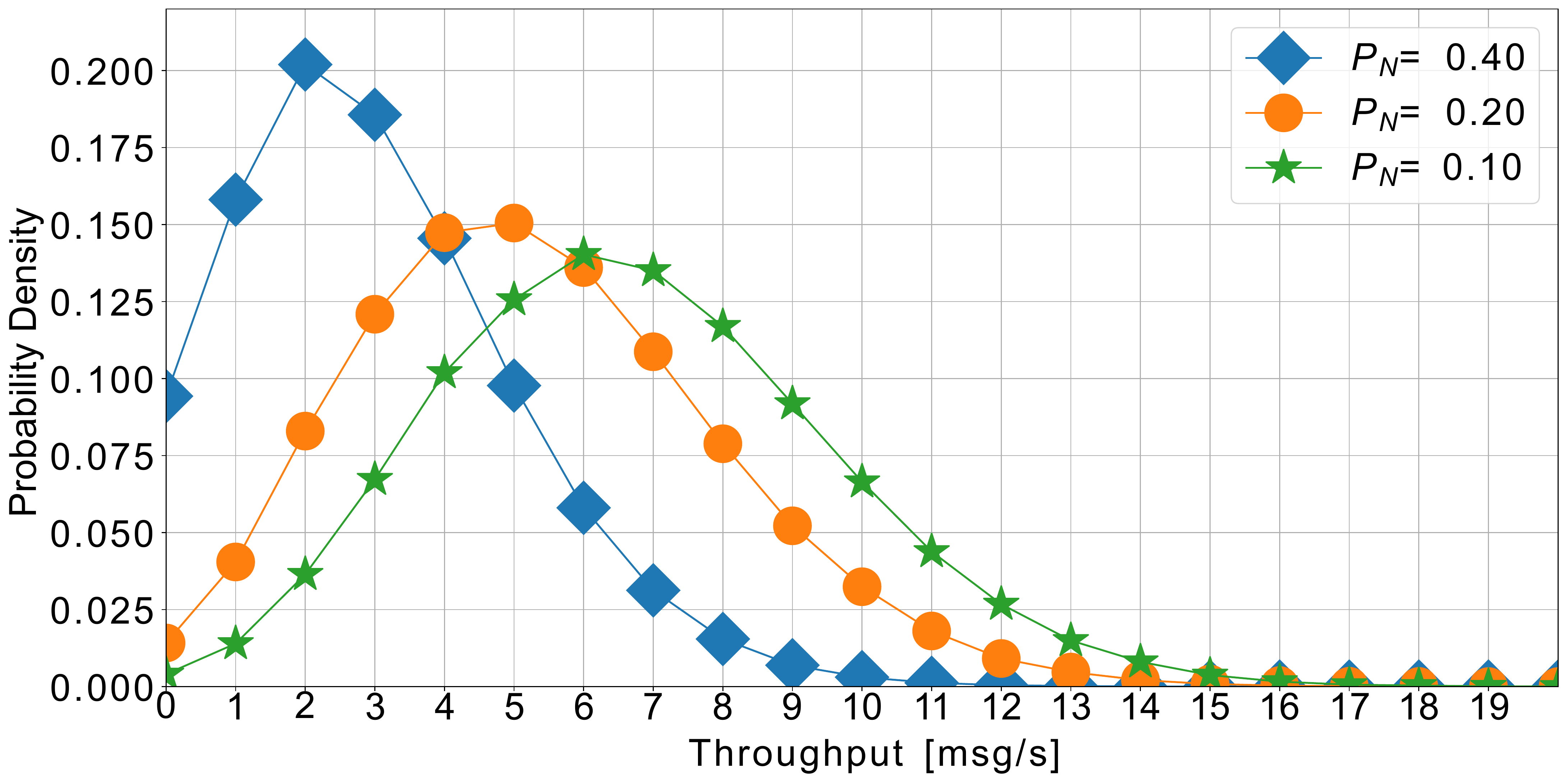}
    \caption{Probability density of receiving k messages in a second. The peak is the average throughput. With $P_S=P_B$ and $T_N=100$~ms, multiple values of $P_N$ have been studied. Increasing $P_N$ reduces the broadcast/scan throughput.}
    \label{fig:PN_tuning}
\end{figure}

In Figure \ref{fig:PN_tuning}, the broadcast - scan throughput is computed for different values of $P_N$. The trade-off between telemetry and broadcast-scan is very important because the primary function of drone's Wi-Fi is telemetry. 
The choice of $T_N$ is very hardware dependent: very small values might cause memory errors, like buffer overflows but large values result in long periods of time when the location update is not possible. 
$T_N=100$~ms is the best trade-off that could be found for the ESP32 hardware used for this research. Using the same values for simulations and experiments allows to compare the results and understand other constraints not included in the behavioral model.

\section{Experimental results}
\label{sec:experimental_results}
In this section, we describe the used hardware and its constraints as well as the experimental campaigns in the laboratory and on the field with real drones. Laboratory experiments are quite important to understand how the real hardware behaves compared to the simulation model.
After the hardware has been validated in the lab, drone experiments could be performed to evaluate how the protocol behaves versus distance in a more realistic scenario as what could be obtained only in ideal conditions.
\subsection{Hardware}\label{sec:wifi_modules}

The Wi-Fi modules used for the experiments are wipy 3.0 by Pycom \cite{pycom_web} and are based upon the ESP32\cite{esp32_docs}.  They are some of the lowest cost Wi-Fi solutions we could find at the time of starting this research, and we have selected them because of their low weight and cost. As a result, we can imagine using them for every drone. 
Each module is equipped with an external antenna to improve the RF performance which otherwise may be hindered by the ceramic chip antenna mounted on the PCB.
The behavioral model is implemented with Arduino (C++) which substitutes the default Micropython environment.

The physical nodes are programmed to behave like the simulated nodes with some small differences dictated by the nature of the device. 

Each module is controlled by its own thread on the PC. 
Since transitions are much less frequent than the period of the system clock, the time accuracy is not a concern and does not invalidate the statistical analysis of the results.
The interface between the Wi-Fi modules and the PC is a Serial-to-USB converter (Silabs CP2102) operating at 115200 kbps.
\subsubsection{Limitations of the setup}

\begin{figure}
    \centering
    \includegraphics[width=0.9\linewidth]{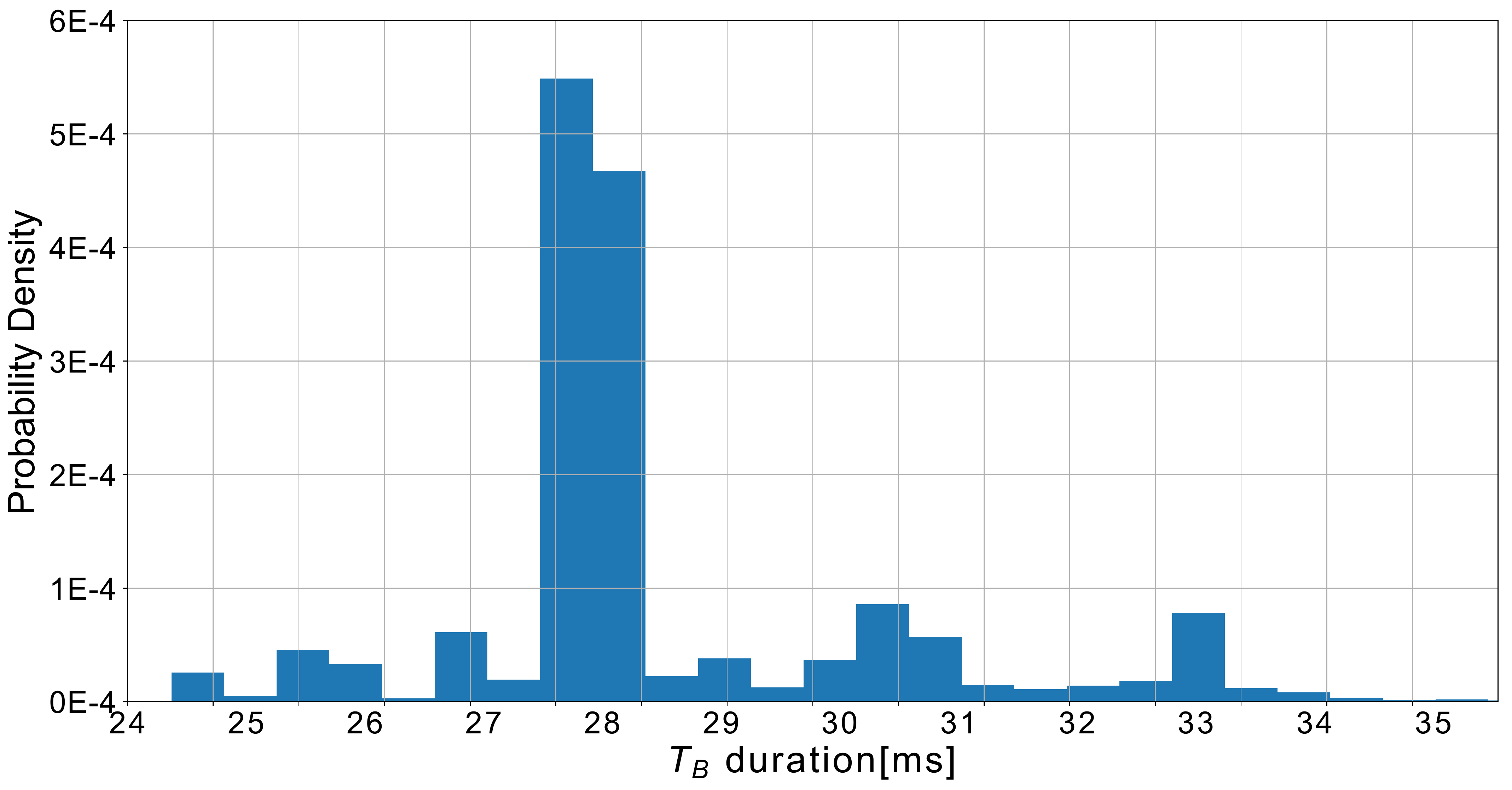}
    \caption{Distribution of $T_{B}$. Although the average is close to 30ms, there is some jitter on $T_{B}$ which can make it vary between 24~ms and 39~ms.}
    \label{fig:TB_dist}
\end{figure}

During our experimental campaign we could observe the presence of jitter on the broadcast interval duration $T_B$. This is because the UART communication is asynchronous and the read operation on the embedded module is blocking with timeout. To mitigate this, it is advisable to enable flow control, depending on the selected serial speed. Up to 230400bps, the so called software flow control (XON/XOFF) is sufficient to ensure correct operation, for faster speeds, hardware flow control (RTS/CTS) needs to be implemented.
The effect of this jitter, whose distribution is shown in Figure \ref{fig:TB_dist}, is noticeable in the results.

It is important to keep in mind the limited amount of memory of these embedded devices. When the required broadcast rate is too high, it is easy to completely fill in the Wi-Fi memory buffer and to prevent any transmission.
The two possible countermeasures are to slow down the transmission rate (e.g. by extending the broadcast time with a short delay() call), or by ensuring that more time is spent scanning than broadcasting.

The scan time $T_S$, which is the time spent listening for incoming transmissions and decoding the messages within them, can be broken into two components: $T_{Rx}$ which is the time spent actually listening to the channel and $T_{comp}$ which is the time spent to extract the received messages from the incoming transmissions and put them in memory in a readable form.
Thus $T_S = T_{comp}+T_{Rx}$.
The amount of messages arriving at the receiver has a considerable impact on $T_{comp}$. 
This translates into a stretch of $T_S$ which is not accompanied by an increase in listening time $T_{Rx}$ but rather into a blind period of duration $T_{comp}$.
While for very low traffic, the approximation $T_S \approx T_{Rx}$ is valid, for higher traffic, $T_{comp}$ must be taken into account.
In simpler words, the scan function takes much longer to execute when more messages are received. 
The additional time seems to be used for processing packets and moving data in memory. 

The extra processing time is due to the fact that the ESP32 is used for something it is not designed for. 
Our educated guess is that, if we could access the low level firmware, the performance can be greatly improved. Espressif provides a proprietary mesh protocol which works up to 32 devices in the same network and its MAC protocol is not very different from beacon exchange. 
We are thus reasonably sure that the hardware is far more capable than what is possible to do just by using its public API. 

\begin{figure}
    \centering
    \includegraphics[width=0.9\linewidth]{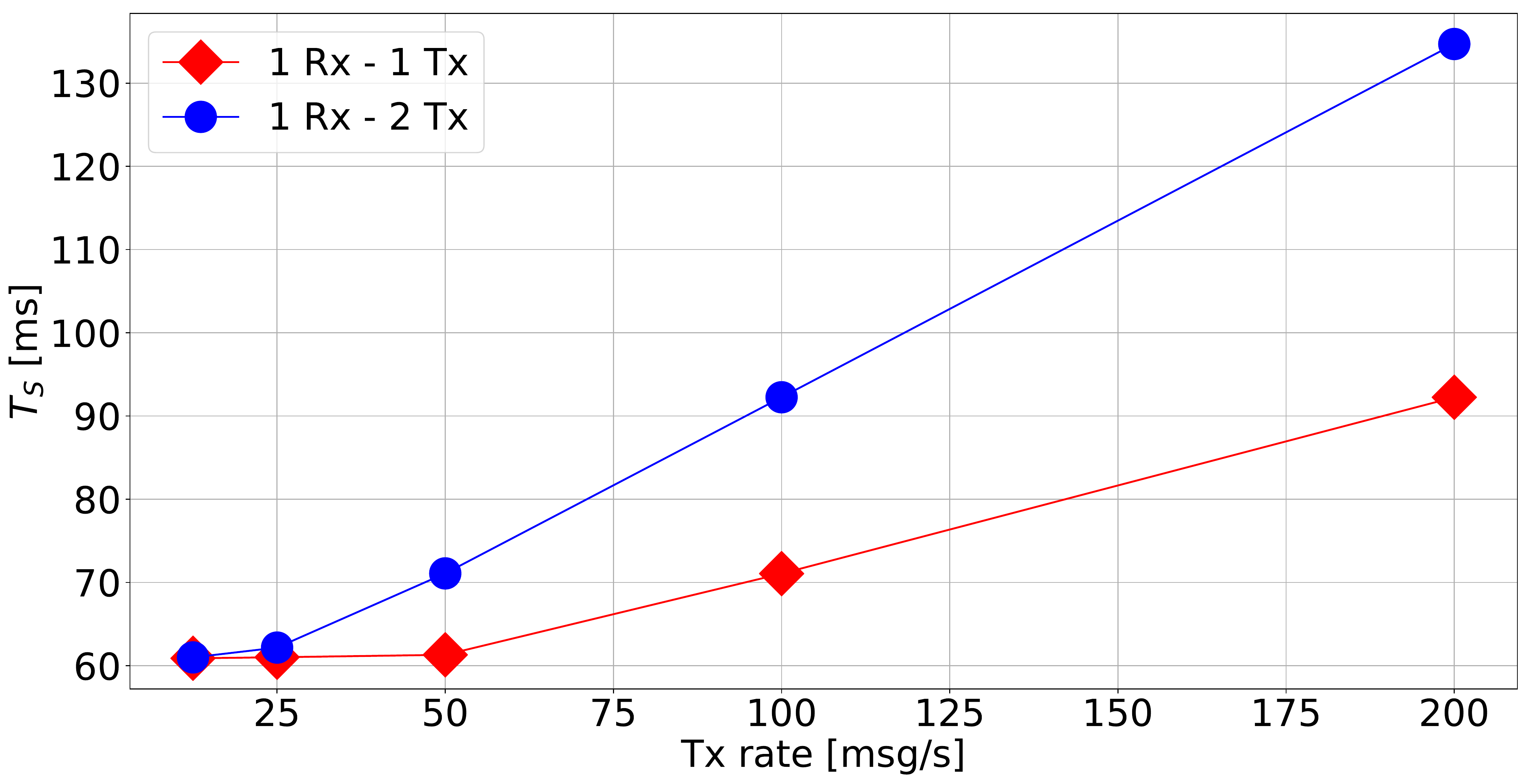}
    \caption{$T_{S}$ vs Transmission rate. An increase in the message transmission rate has a big impact on the processing time. This behavior is dominant over radio collisions. In fact, the time spent processing the incoming messages becomes higher than the time spent receiving them.}
    \label{fig:trx_vs_txrate}
\end{figure}

\begin{figure}
    \centering
    \includegraphics[width=0.9\linewidth]{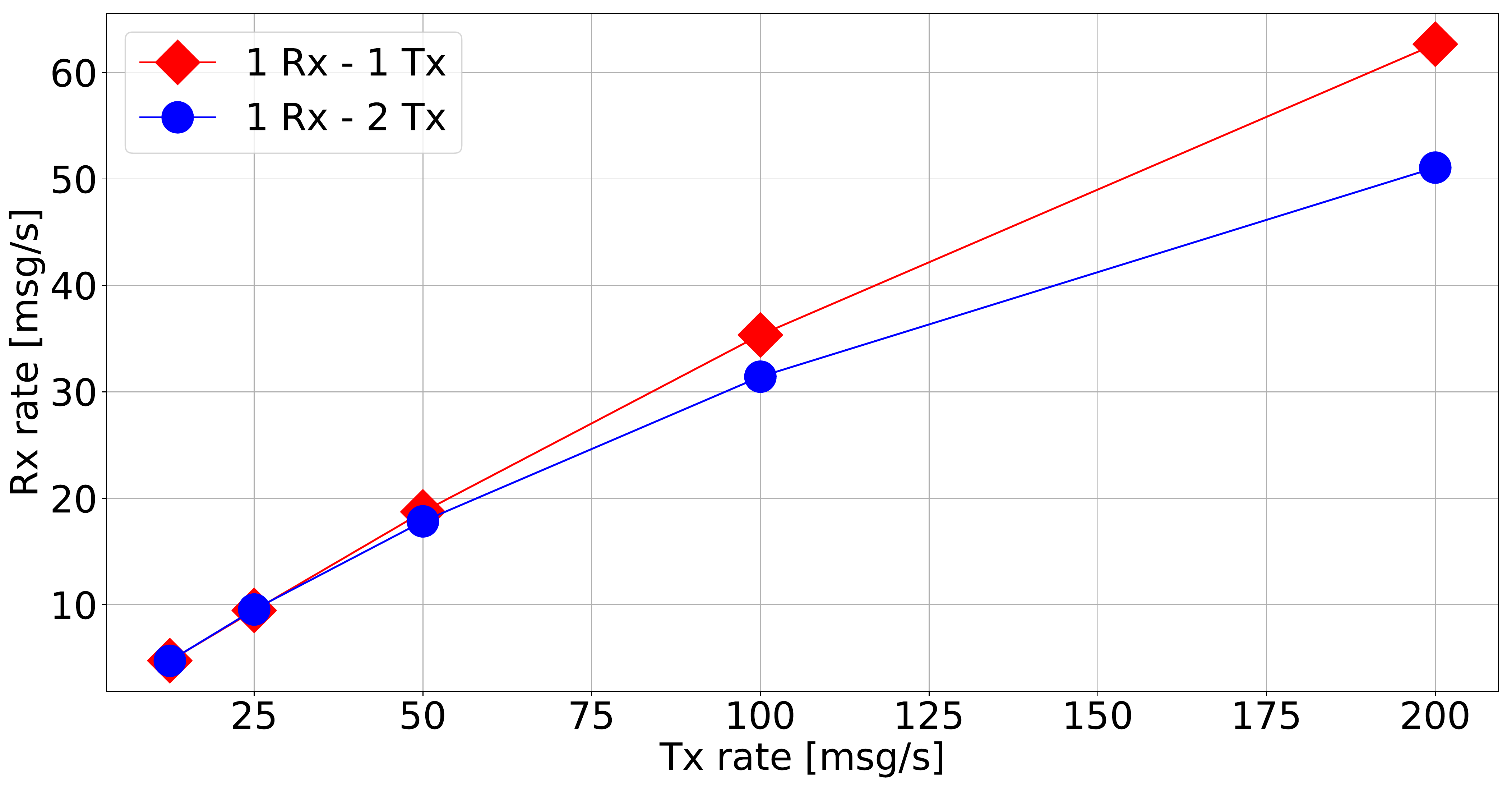}
    \caption{Reception rate vs Transmission rate. The increase in $T_{S}$ causes the receiver throughput to have a reduced slope compared to the ideal linear case.}
    \label{fig:rx_rate_vs_txrate}
\end{figure}

Figures \ref{fig:trx_vs_txrate} and \ref{fig:rx_rate_vs_txrate} show the impact of the transmission rate on the receive time and the reception rate respectively. 
It is clear that the more messages are transmitted, the longer it takes to the wi-fi chip to process them impacting negatively the reception rate.
While the transmission rate has barely any impact below 50~msg/s, the receiver struggles progressively with the increase of incoming messages to be processed.
With two transmitters, the turning point happens even earlier because the rate of incoming messages is the double. 
Figure \ref{fig:rx_rate_vs_txrate} shows how the reception rate does not follow a linear trend and its slope reduces more and more with the amount of transmitted messages.

\subsection{Laboratory experiments setup}

In order to get a baseline for the real - world experiments performed with the drones, a complete measurement campaign has been performed in our labs. 
The measurements have been performed both in an office environment and in a Faraday cage. The modules have been programmed to operate according to our behavioral simulation so that results could be compared between simulation and experiments.
The real modules, programmed with the same parameters as in Table \ref{tab:sim_timingl}, give very similar results as our simulation model but also show how perfection does not belong to our world. 
The physical limitations and constraints of the ESP32 are discussed in detail in Section \ref{sec:wifi_modules}.

The modules have been placed approximately 50~cm away from each other to minimise propagation effects.

The time spent for each operation has been measured on the ESP32 Wi-Fi chip both with an oscilloscope, by toggling some of the pins at the beginning and end of the timed functions, and with the internal timer of the ESP32.

Each device is connected via serial port to a Raspberry pi which saves all the incoming beacons together with its timestamp. 
The time is measured from the beginning of the logging program.
In the simulations, we assume that all the drones broadcast beacons, scan for beacons and do networking operations with the same probability as any other drone. In other words, every drone uses the same protocol parameters.

From the probabilistic model, we can derive the fact that $P_S$ is the limiting factor in the amount of messages that can be received. 

The higher $P_S$, the more more messages get received in proportion to the transmitted ones. 
However, a higher $P_S$ means a lower $P_B$ which implies that less messages are transmitted.
The maximum throughput is only obtainable when $P_S = P_B$ and, when no networking is present, at most a half of the transmitted messages can be received by a single drone.

Both experiments and analysis show that, as expected, the delay between successfully received packets follows an exponential distribution (see Figure \ref{fig:tdoa}).
 
Since messages can be lost due to many causes (e.g. congestion, multipath fading, bad antenna alignment), it is important to maximize the throughput to share always the most updated position available. The maximum throughput can be obtained by having $P_B=P_S=0.5$;

It is also worth noting that a low throughput results in a flight speed limitation.

\subsection{Laboratory experiments results}

\subsubsection{Broadcast - Scan}

\begin{figure}
    \centering
    \includegraphics[width=0.98\linewidth]{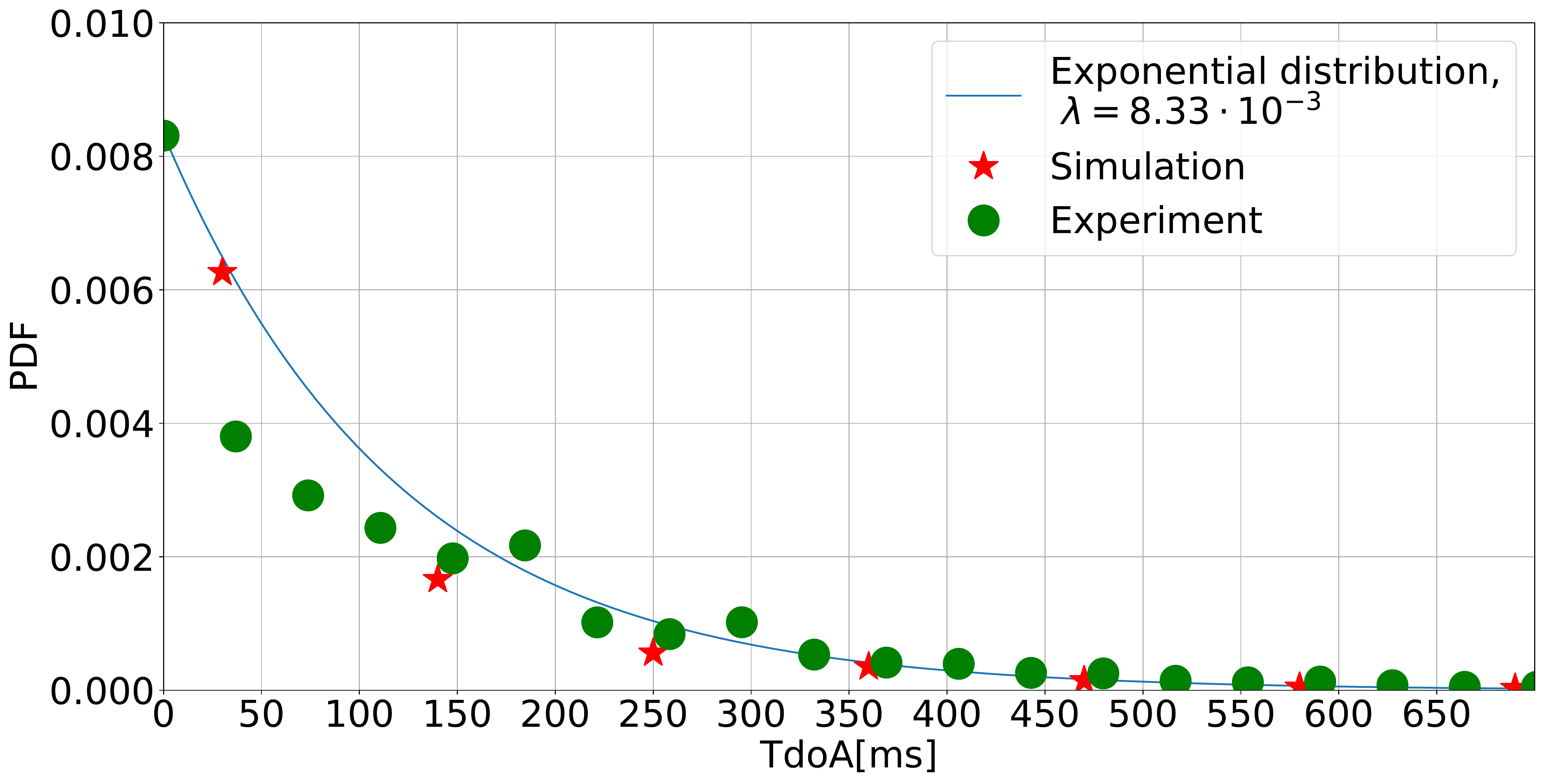}
    \caption{Distribution of time of arrival of the messages: simulation vs experimental vs $\lambda e^{-\lambda x}$ with $\lambda = 1/2T_{S} = 0.00833$, $P_S = P_B = 0.5$, and $T_S$ and $T_B$ as reported in table \ref{tab:sim_timingl}.}
     \label{fig:tdoa}
\end{figure}

The simplest implementation of the communication protocol involves only broadcast and scan, without networking. The behavior of each module can then be modeled as a Poisson process. 
The time of arrival follows an exponential distribution (see Figure \ref{fig:tdoa}) which validates this assumption.

\begin{figure}
    \centering
    \includegraphics[width=0.98\linewidth]{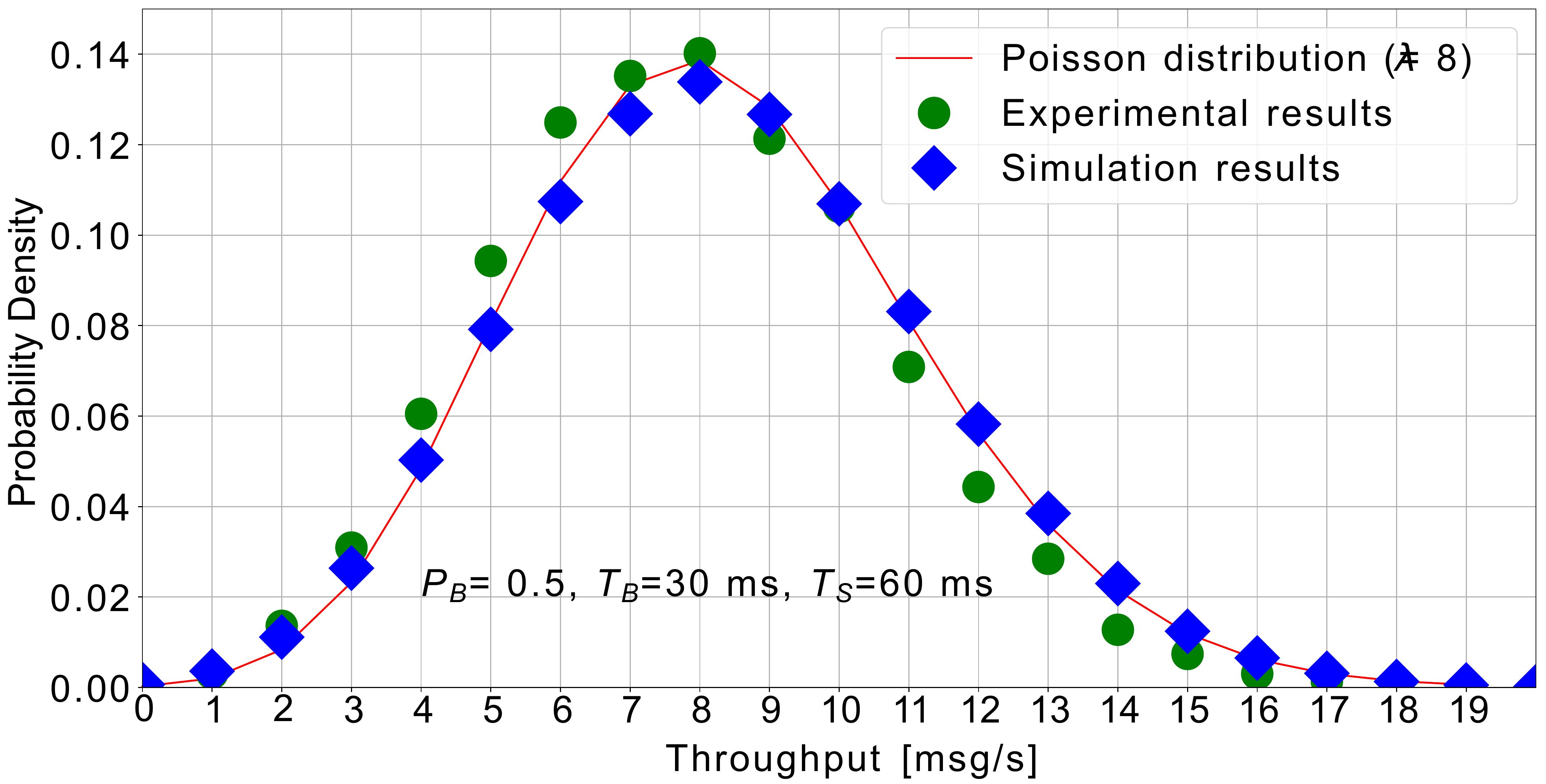}
    \caption{Probability density of receiving k messages in a second: Simulation vs Experimental results for $P_B=P_S=0.5$.}
    \label{fig:exp_vs_simulation}
\end{figure}

The experimental results are in agreement with the simulations results.  However, the physical device shows some limitations and constraints which were not considered in the simulations. 
Despite the device limitations, the results are very close to what could be observed by simulating the behavioral model.

With $P_B = P_S = 0.5$ and the settings of Table \ref{tab:sim_timingl}, the average throughput is slightly higher than 8~msg/s which is half the broadcast rate. 
This has been confirmed by both simulations and experiments in the laboratory (see Figure \ref{fig:exp_vs_simulation}) and it is in accordance with our mathematical model. 
For example, the number of successful receptions in a 1 s window follows \eqref{eq:n_succ}, resulting in $\overline{{N}_{success}} = 8$. 
As shown in Figures \ref{fig:tdoa} and \ref{fig:exp_vs_simulation}, the results match very well both simulations and experiments with the real devices.

\subsubsection{Radio Collisions}
\label{subsec:collisions}
\begin{figure}
    \centering
    \includegraphics[width=0.95\linewidth]{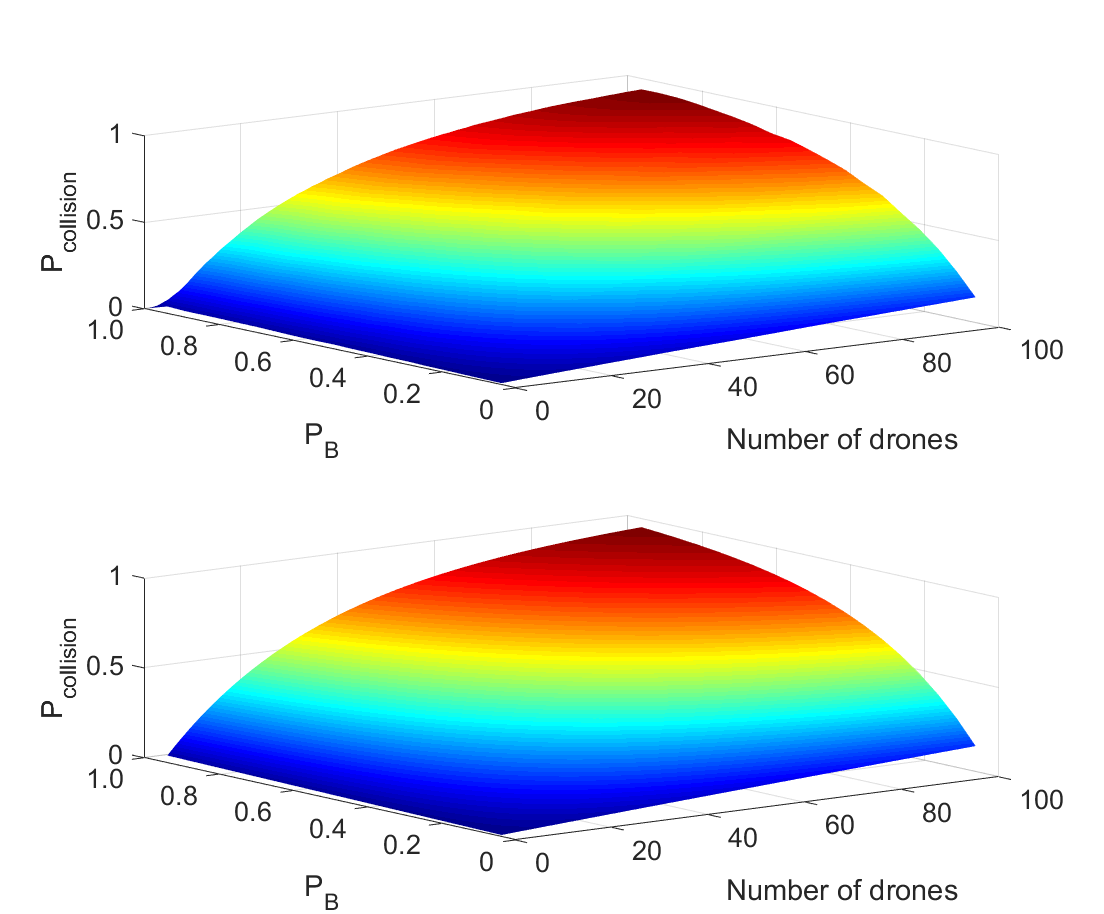}
    \caption{Cumulative distribution function of the probability of collisions depending on the amount of drones (0 to 100) and the probability of broadcasting (0 to 1). 
The bottom image is the theoretical result(see \eqref{eq:P_collision_kdrones}), the top image is the simulation result.}
    \label{fig:collisions}
\end{figure}

The amount of information needed to avoid accidents when multiple drones are close to each other is much higher than when the drones are far apart. 
Thus, the effect of radio collisions needs to be carefully evaluated in order to predict the amount of drones that can safely fly in the same area.
Moreover, it is important to consider the effect of hardware "saturation" explained in Section \ref{sec:wifi_modules}.

Figure \ref{fig:collisions} shows the probability of collision between transmissions depending on the amount of active drones and the probability that they are in the broadcast mode. 
The simulation result matches closely the theoretical model of equation  \eqref{eq:P_collision_kdrones}. 
In theory, the effect of radio collisions is small on the system performance because the message length is very short compared to the time between consecutive transmissions (less than 1~ms vs tens or hundreds of ms).

However, experiments with the real devices show that the limiting factor is the hardware. 
As mentioned in Section \ref{sec:wifi_modules}, the processing time of received packets increases proportionally to the number of incoming messages to the point where it is not negligible anymore.

However, this is somehow a device specific behavior which can be mitigated by both hardware and firmware adjustments (e.g. faster hardware, bigger receive buffer, algorithmic optimizations, faster communication interface).

Summarizing, we observed from our experiments that processing time increase is dominant over radio collisions.

\subsubsection{Broadcast, Scan, Networking}
\begin{figure}
    \centering
    \includegraphics[width=0.98\linewidth]{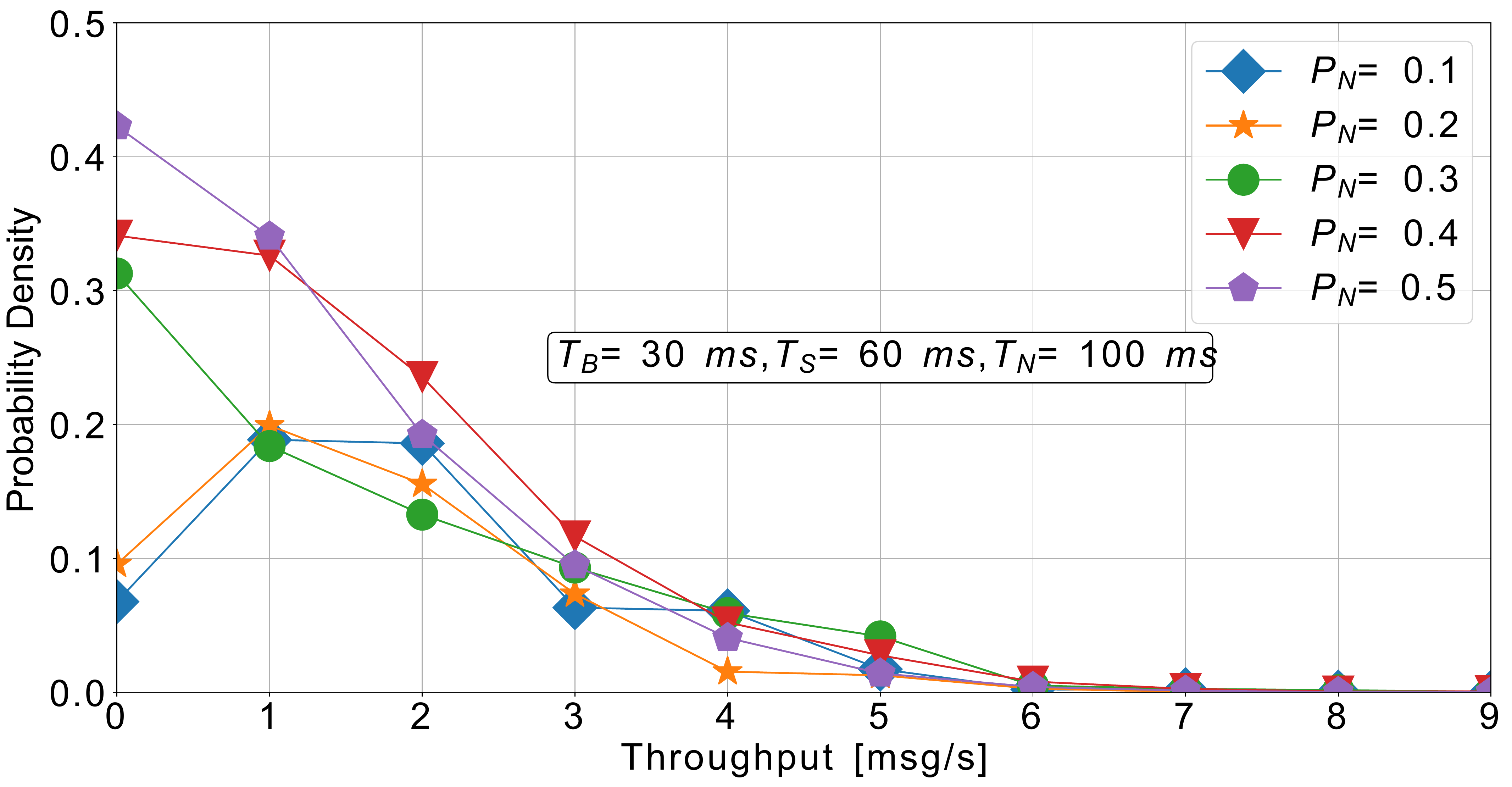}
    \caption{Probability density (experimental results) of receiving k messages in a second for different networking rates. A trade off must be chosen between telemetry and  broadcast-scan in order to guarantee a minimum throughput of 1~msg/s.}
    \label{fig:rx_dist_net}
\end{figure}

\begin{figure}
    \centering
    \includegraphics[width=0.98\linewidth]{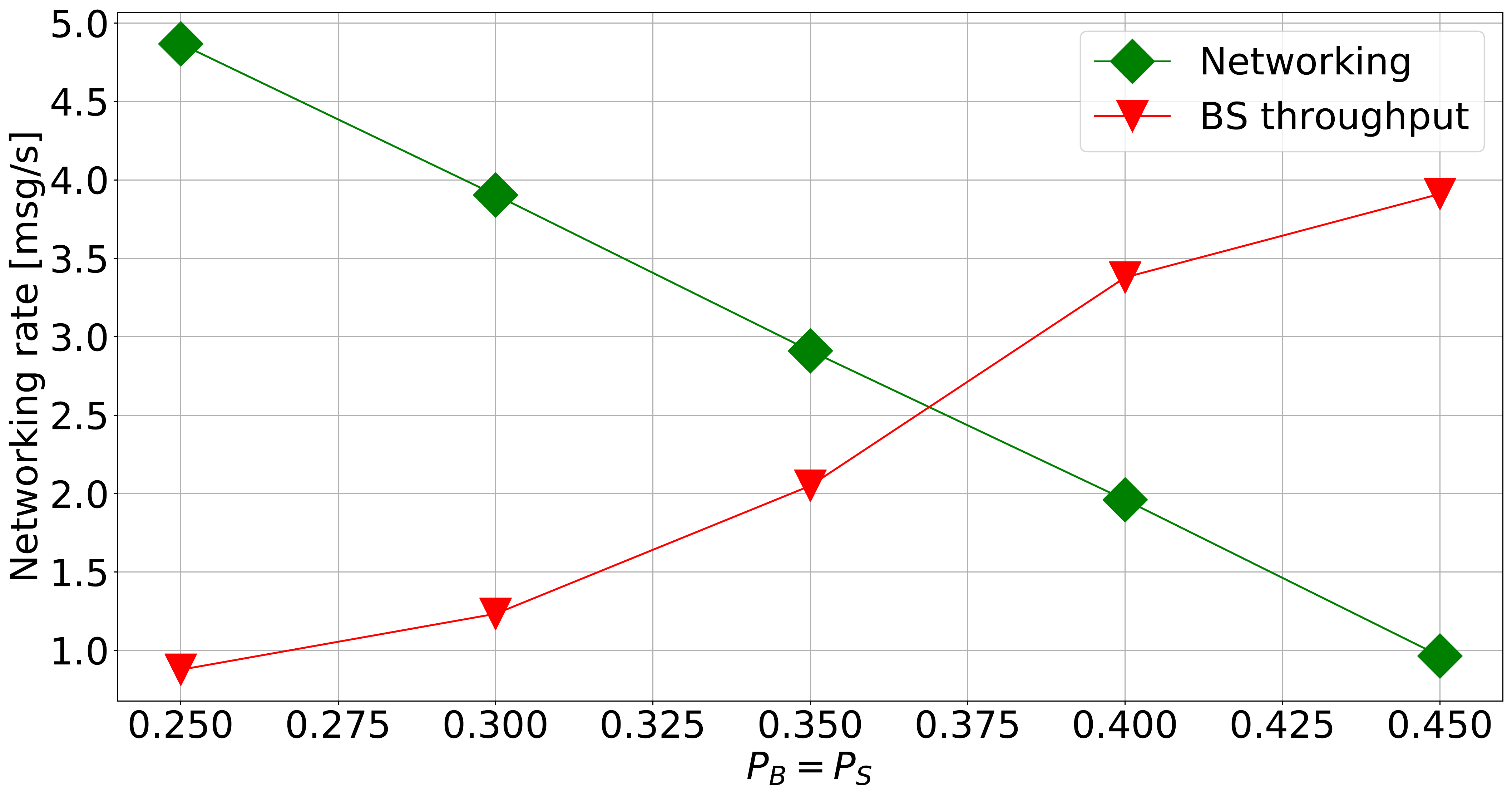}
    \caption{Networking rate and Broadcast - Scan Throughput vs $P_B=P_S$.
As intuition suggests, the more time is spent broadcasting and scanning, the less time can be spent for networking and telemetry. 
Depending on the application it is necessary to find  a mission dependent trade off between telemetry, which is typically the primary function of Wi-Fi in small drones, and broadcast scan.}
    \label{fig:net_rate_vs_pb}
\end{figure}

Introducing networking periods is going to impact the performance of the broadcast - scan mechanism. 
In particular, we noticed how the broadcast and scan rates must be reduced to accommodate for the processing of network packets.
From a conflict management perspective, the parameters should be set in order to move the curve of Figure \ref{fig:rx_dist_net} to the right until the probability of getting 0 msg/s is 0 but this would greatly reduce the network throughput.
It is important at the design stage to find a proper trade-off between network/telemetry and broadcast-scan such that at least 1~msg/s can be successfully exchanged without disrupting the telemetry link.  
According to Table 14 in \cite{itu_r_report}, the requirement for control and navigation aid data rate can go to approximately 12.2~kbps, which poses a minimum requirement for the network throughput of at least 1~packet/s, since Wi-Fi packets can be up to approximately 12~kbit in length. 
The experimental results are shown in Figures \ref{fig:rx_dist_net} and \ref{fig:net_rate_vs_pb}. 
As expected, the more time is allocated to the beacon broadcasting, the higher the probability that messages get received at the expense of the network throughput. 

\subsection{Drone experiments}
\begin{figure}
    \centering
    \includegraphics[width=0.98\linewidth]{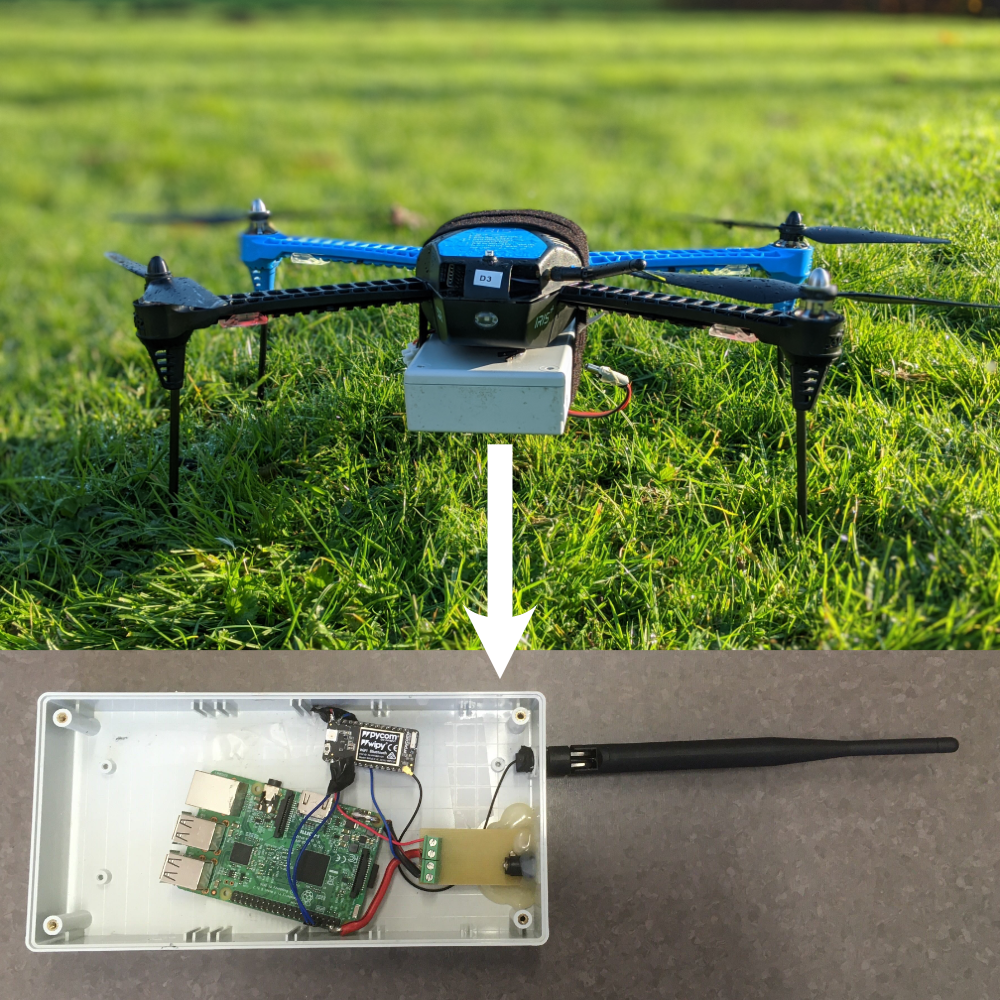}
    \caption{Setup for the drone experiments. Top: fully assembled drone (3DR Iris+) with attached the CAP box. 
Bottom: the open box containing a Raspberry pi, the ESP32 module and a small power board to convert the 12 V from the battery to 5 V.}
    \label{fig:drone_setup}
\end{figure}
We also study Wi-Fi coverage on a drone. 
The system performance has been tested at different distances to verify up to which point the Wi-Fi is still useful for conflict management.
The experiments have been conducted using two identical devices called Collision Avoidance Packages (CAPs). Each CAP includes a Raspberry Pi 3B+, a GPS module and a ESP32 equipped with a stylus antenna. 
The setup is shown in Figure \ref{fig:drone_setup}.
The Raspberry Pi is responsible for reading the GPS coordinates, encoding them and passing them to the ESP32 via serial port. 
The Pi also collects the incoming data from the ESP32 and saves them on a text file.
When the CAP is connected to a drone, it acts as companion computer for the flight computer, giving high level commands such as take off, land or go to a way-point, and it is responsible for running the physical collision avoidance algorithm.
For this research work, though, no physical collision avoidance has been activated since the purpose has been to characterize the communication system.

One of the CAPs was on the highest tower of our department, in order to simulate a drone deployment. 
A second CAP was connected to a drone hovering at the same altitude as the tower but at different positions in the Arenberg Campus (See Figure \ref{fig:los_nlos}).
This setup was used to characterize the communication performance vs distance. 
The node positioned on top of our department tower is composed by the exact same hardware and software as the package mounted on the UAV.

\subsection{Drone results}
\begin{figure}
\label{fig:los_nlos}
  \centering \includegraphics[width=1\columnwidth]{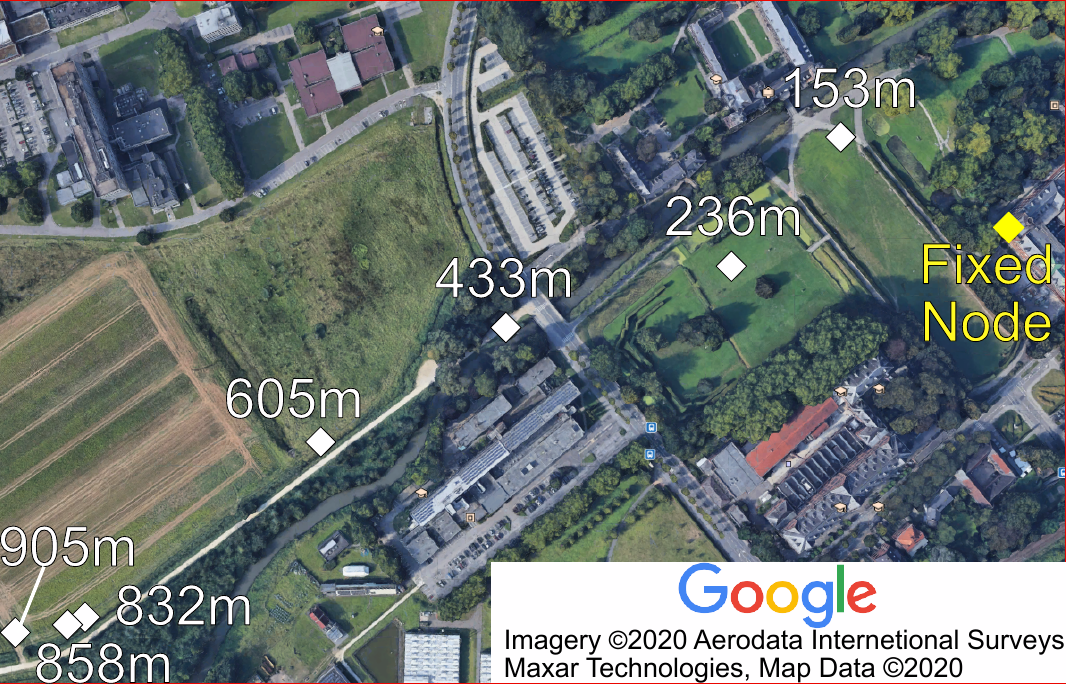}
  \caption{Map of the measurement points}
\end{figure}

\begin{figure}
    \centering
    \includegraphics[width=0.98\linewidth]{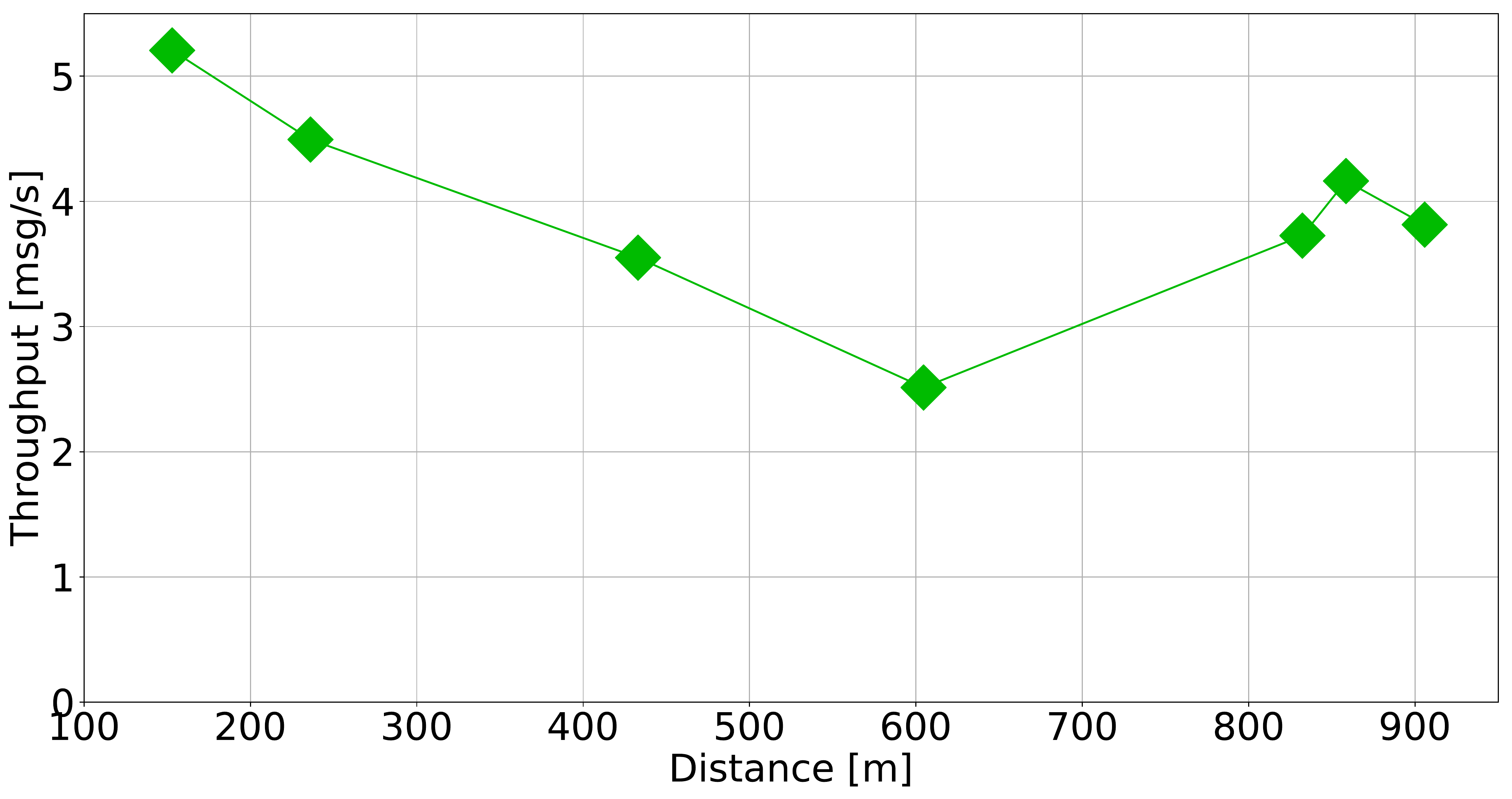}
    \caption{Throughput vs distance. The area between 400m and 600m is covered in vegetation and trees and next to a set of tall buildings. This affects negatively the RSSI performance. The last two points, although not line of sight, are taken in the middle of an open field with almost no obstacles in the Fresnel zone. }
    \label{fig:rate_vs_distance}
\end{figure}
\begin{figure}
    \centering
    \includegraphics[width=0.98\linewidth]{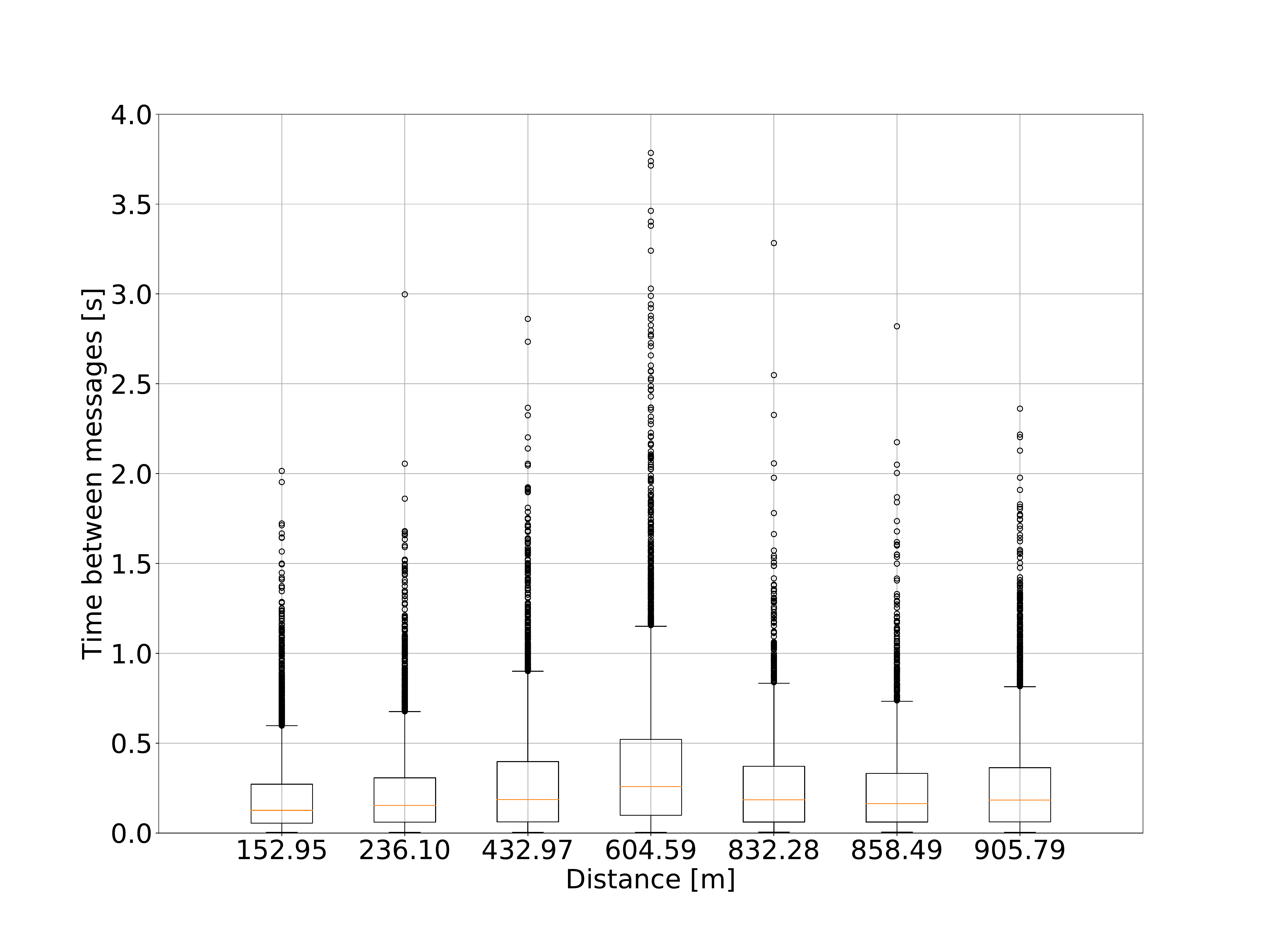}
    \caption{This figure shows the time between consecutive messages at difference distances. Where vegetation and buildings are more dense, the recorded time values have a higher dispersion as it is also reflected by the average throughput.}
    \label{fig:tdoa_vs_distance}
\end{figure}

The drone experiments have been performed in the Arenberg campus in Heverlee, Belgium (see Figure \ref{fig:los_nlos}.
The environment can be considered sub-urban with high vegetation and trees surrounding sparse tall buildings.
The furthest measurement point from the tower is approximately 900~m far.

Even at the furthest location, it was possible to receive approximately 4 msg/s with low Received Signal Strength Indicator (RSSI, -85~dBm on average). 

\subsubsection{RSSI vs Distance}
\begin{figure}
    \centering
    \includegraphics[width=0.98\linewidth]{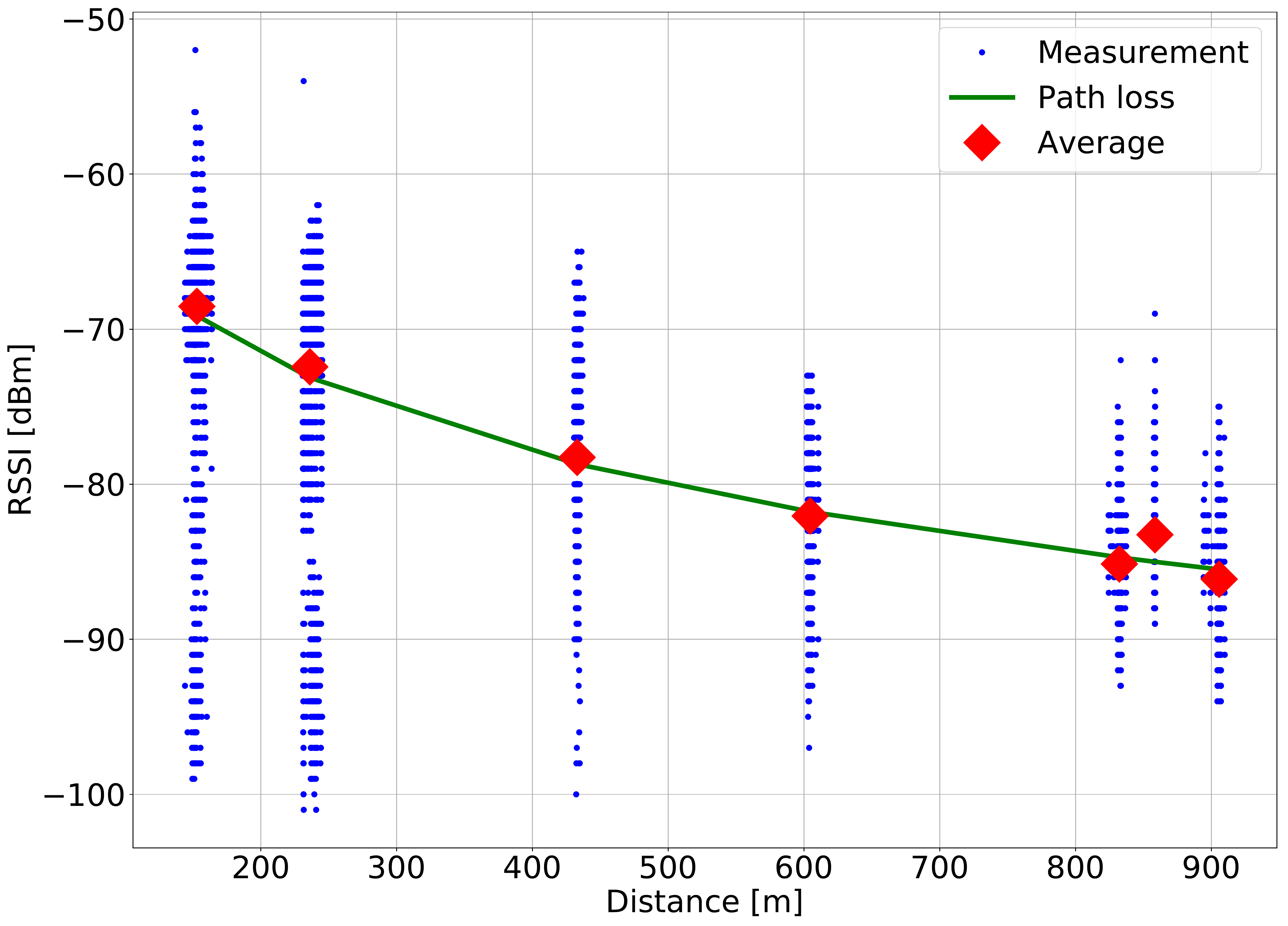}
    \caption{The RSSI, follows pretty well the free space path loss, even though a few points are not in line of sight. The path loss parameters are: $P_{t}=19.5$~dBm, $K=\lambda/(4\pi d_{0})=-3.55$, $d_{0}=2D/\lambda = 0.0147$, $\gamma=2.118$. Towards the end, there is a slight increase in the average RSSI which depends from receiving only the messages with better RSSI and losing the messages with lower RSSI due to low SNR.}
    \label{fig:rssi_vs_distance}
\end{figure}

As shown in Figure \ref{fig:rssi_vs_distance}, our RSSI measurements follow pretty well the log distance path loss model $P_{r} = P_{t}-K-10\gamma log_{10}(d/d_{0})$, where:
\begin{itemize}
    \item $P_{r}$ is the received power;
    \item $P_{t}=19.5$~dBm is the transmit power;
    \item $d_{0}=2D/\lambda = 0.0147$ is the far field distance for the Wi-Fi middle frequency of 2.45 GHz;
    \item $K=\lambda/(4\pi d_{0})=3.55$ is the path loss factor;
    \item $\gamma=2.118$ is the path loss exponent extrapolated from the measurements.
\end{itemize}
At close distance, it still possible to receive beacons arriving with lower power, that is why the bottom part of the plot is more populated at distances lower than 500~m than it is for further distances.
This result is encouraging as it shows that messages can still be received at 900~m distance. 
However, signal strengths below -80~dBm indicate the need for a different radio system for distances greater than 1~km.

\subsubsection{Throughput vs Distance}

The throughput suffered for the presence of buildings and vegetation (see Figures \ref{fig:rate_vs_distance} and \ref{fig:tdoa_vs_distance}) dropping to 2.5~msg/s but the system overall remained functional. 
This might be an important factor to keep into account for missions in dense urban or forest environments. 
Another important lesson to bring home is that correctly positioning the antennas is very important. 
The UAV frame itself can cause shadowing to the antennas and reduce the communication performance \cite{vinogradov2018tutorial}\cite{comp_survey_channel_modeling_uav}.
\section{Conclusions}

In this paper, we propose a method for exploiting the Wi-Fi modules already present in many small UAVs as a tool for sense and avoid.
The main concept is to use Wi-Fi beacons as a mechanism to broadcast positional information of the UAVs.

Our simulation results show that the system can be tuned to have good reliability and it is able to deliver more than one message per second, even when a considerable number of drones share the same radio channel.

The experimental results obtained  with low cost Wi-Fi devices agree perfectly with the simulation results.

Measurements performed with drones also show that the system works reliably even at distances close to 900~m, leaving enough room to the UAV to safely perform an avoidance maneuver.

\bibliographystyle{plain}
\bibliography{bibliography}
\end{document}